\documentclass[manuscript]{acmart}

\AtBeginDocument{%
  }

\setcopyright{none}
\copyrightyear{2026}
\acmYear{2026}
\acmDOI{XXXXXXX.XXXXXXX}
\acmJournal{TSC}

\usepackage{graphicx}

\usepackage{booktabs}
\newcommand{\descr}[1]{\smallskip\noindent\textbf{#1}}
\usepackage{subcaption}
\usepackage{siunitx}  %
\usepackage{mdframed}
\usepackage{adjustbox}
\usepackage{xcolor}
\usepackage{tcolorbox}
\usepackage{amsmath}
\usepackage{subcaption}
\usepackage{siunitx}
\usepackage{multirow}
\usepackage{float}
\usepackage{longtable}
\usepackage{tabularx}
\usepackage{seqsplit}
\usepackage{tikz}
\usetikzlibrary{positioning}

\newif\ifcomment
\commenttrue
\ifcomment
\newcommand{\ms}[1]{{\bf \textcolor{blue}{MS: #1}}}
\newcommand{\jbnote}[1]{{\bf \textcolor{magenta}{JB: #1}}}
\newcommand{\ub}[1]{{\bf \textcolor{orange}{UB: #1}}}
\else
\newcommand{\ms}[1]{}
\newcommand{\jbnote}[1]{}
\newcommand{\ub}[1]{}
\fi

\usepackage{colortbl}
\definecolor{lightteal}{HTML}{A1DADC}
\definecolor{lightorange}{HTML}{F7C79E}
\newif\ifhighlight
\highlighttrue
\ifhighlight

\else

\fi

\begin{document}
\author{Utkucan Balci}
\affiliation{%
  \institution{Binghamton University}
  \country{United States}
}

\author{Michael Sirivianos}
\affiliation{%
  \institution{Cyprus University of Technology}
  \country{Cyprus}
}
\email{michael.sirivianos@cut.ac.cy}

\author{Jeremy Blackburn}
\affiliation{%
  \institution{Binghamton University}
  \country{United States}
}
\email{jblackbu@binghamton.edu }

\title{Roll in the Tanks! Measuring Left-wing Extremism on Reddit at Scale}

\begin{abstract}
      Social media's role in the spread and evolution of extremism is a focus of intense study.
      Online extremists have been involved in the dissemination of online hate, mis- and disinformation, and real-world violence.
      While the majority of research has focused on \emph{right-wing} extremism, recent real-world incidents have highlighted the potential for far-left extremists to engage in violence and cause real-world harm as well.
      In this paper, we present the first large-scale measurement of \emph{left-wing} extremism on social media.
      Analyzing 1.3 million posts from 53,000 authors from tankie subreddits, we focus on ``tankies,'' a left-wing community that first arose in the 1950s in support of hardline actions of the USSR and has evolved to support what they call ``Actually Existing Socialist'' countries, e.g., CCP-run China, the USSR, and North Korea.
      Among other things, our analysis reveals that these groups occupy the periphery of the broader far-left community on Reddit, and their discourse distinctively focus on state-level politics and support for authoritarian regimes, rather than on social justice issues.
      Finally, we show that tankies have high toxicity scores and use pejorative language, mirroring toxicity patterns reported for other online extremist communities.
      Our findings provide empirical evidence of the distinct positioning and discourse of left-wing extremist groups on social media.

    \end{abstract}

\maketitle

    \section{Introduction}
    \label{sec:intro}
    
    The use of social media by extremists is well documented in the press~\cite{sands2021,theguardian2021,bump2022} and has been a heavy focus of the research community~\cite{aldera2021online,mamie2021anti,grover2019detecting}.
    However, almost all recent work has studied \emph{right-wing} extremists.
    In large part, this is due to the recent rise of right-wing populist parties and leaders globally~\cite{rooduijn2019state}, which in turn has led to an abundance of easily identifiable right-wing extremists online that are relatively prolific when it comes to content production.
    As a result, these measurement studies have yielded deep insights into right-wing extremist behaviors, network structures, and influence across social media platforms~\cite{zannettou2018origins,hine2017kek, zannettou2017web}.

    At the same time, there has been a steady rise in political rhetoric characterizing mainstream political parties as far-left extremists, scapegoating the far-left for violent activities.
    Examples include claims that antifascist activists orchestrated the January 6th Insurrection~\cite{beer2021}, allegations that left-wing extremists were responsible for violence during the George Floyd protests~\cite{devine2020}, or even accusations of left-wing groups setting forest fires in Oregon~\cite{harris2020}. 
    Much of this rhetoric can be dismissed as political posturing, given that mainstream liberal or left-of-center figures do not fit any reasonable definition of far-left extremism (e.g., the 46th President of the United States, Joe Biden, does not fit any definition of far-left that we have found in the literature).

    However, real-world events demonstrate that authoritarian-oriented communities on the far-left also present tangible risks that go beyond political rhetoric.
    For example, in May 2025, two Israeli embassy staff members were killed outside the Capital Jewish Museum in Washington, D.C., in an attack that investigators linked to far-left extremists~\cite{Yousef2025NPR}.
    In another example, a son of a CIA deputy director was influenced by pro-authoritarian and anti-Western narratives circulating online, ultimately joining Russian forces in Ukraine, where he was killed in combat~\cite{DeLuceNBCNews}. 
    These cases highlight that online spaces associated with far-left extremists are not just a theoretical concern, but can also be a source of real-world recruitment and violence.

    \begin{figure*}[t]
      \centering
      \includegraphics[width=\textwidth]{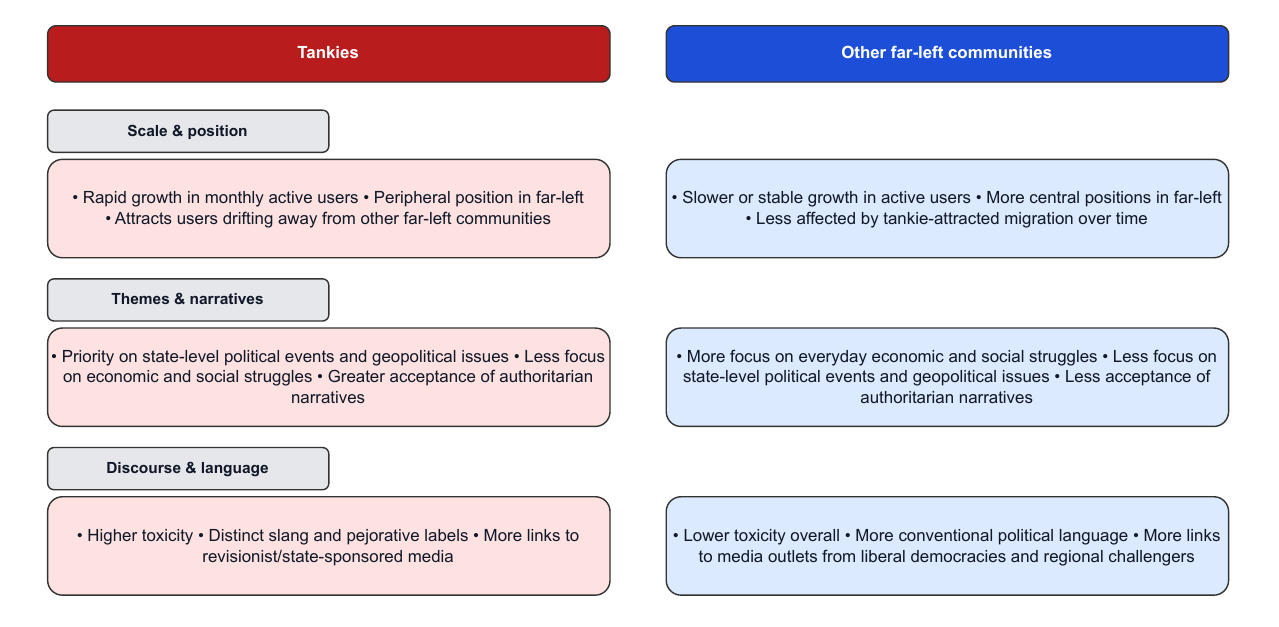}
      \caption{Conceptual overview of how tankies differ from other far-left communities on Reddit.}
      \label{fig:conceptual}
  \end{figure*}

    Despite such incidents, and a history of violence and chaos attributed to far-left extremists, there are essentially no large-scale, data-driven studies of the far-left on social media, let alone far-left extremists.
    This lack of empirical research leaves a gap in our understanding of \emph{whether} and \emph{how} far-left extremism manifests online.
    
    In this paper, we address this gap by conducting the first large-scale measurements of the far-left on social media, using a variety of measurement techniques.
    Following March~\cite{march2008contemporary}, we use \emph{far-left} to refer to actors that place themselves to the left of social democracy and advocate a fundamental transformation of capitalist societies.
    We focus primarily on a large left-wing community known as \emph{tankies}.
    In our analysis, we treat tankies as one specific current within this broader far-left ecosystem. 
    We compare their networks and discourse to other prominent far-left communities to identify when, and in what respects, they adopt more extreme positions and behaviors.
    Historically, tankies were supporters of hard-line Soviet actions~\cite{glastonbury1998children}; more Stalinist than Leninist.
    The name originates from Soviets using tanks to put down rebellions in eastern Europe~\cite{robinson2011new}.
    More recently, tankies have grown to support the actions of the CCP in China, a currently operational ``Actually Existing Socialist'' (AES) country.
    Notably, their support extends beyond just AES countries, often siding with or excusing anti-NATO, non-socialist, autocratic regimes, including Putin-controlled Russia~\cite{carey2022dont}.
    Regardless of their historical ideology, tankies have recently shown behavior similar to the right-wing extremists (e.g., denying the Uyghur genocide~\cite{roberts2020war}).
    Given the parallels in behavior and the lack of measurement studies on left-wing extremism, it is crucial to quantitatively analyze these communities to understand their prevalence and impact on social media platforms.

  \begin{figure*}[t]
    \centering
    \includegraphics[width=\textwidth,height=0.2\textheight]{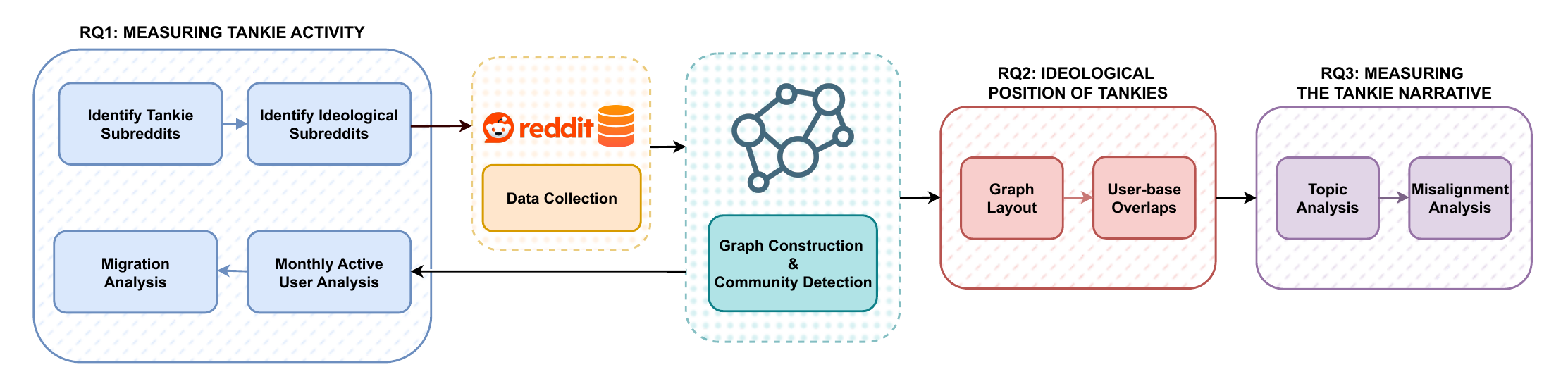}
    \caption{Overview of the data collection and analysis pipeline for measuring tankies on Reddit.}
    \label{fig:diagram}
\end{figure*}

    To this end, we set out to answer three research questions:
    \begin{itemize}
      \item \textbf{RQ1:} What is the scale and growth trajectory of tankie communities on Reddit?
      As a relatively unknown community in the academic literature, the first step to advancing our understanding is to identify the online tankie community and measure their community growth. %
      Establishing how large, active, and fast-growing these communities are is a necessary baseline for assessing whether they are a marginal niche or have a substantial extremist presence on Reddit.

      \item \textbf{RQ2:} Where are tankies positioned in the greater left-wing community?
      While tankies have been characterized in the media and the existing literature as on the extreme end of the left-wing spectrum, at the time of this writing, there is no quantitative evidence to support this positioning.
      By locating tankies within a network of ideological subreddits on Reddit built from shared user participation, we can determine whether they occupy a peripheral position that reflects ideological isolation from the rest of the far-left.
      Showing that tankies are structurally isolated in this way provides quantitative support for treating them as a distinct left-wing extremist community rather than simply another far-left subreddit.
      \item \textbf{RQ3:} What do tankies discuss and how does it differ from other far-left communities?
      We aim to analyze and measure the tankie discourse, comparing it with other far-left communities to better understand its unique characteristics.
      This allows us to determine whether tankies simply intensify themes common to the far-left or instead prioritize distinct issues and actors associated with authoritarian, geopolitically aligned extremism. 
    \end{itemize}

    Figure~\ref{fig:conceptual} depicts a conceptual overview for tankies based on our main empirical findings and compares tankies to other far-left communities. 
    It highlighs how they differ in their relative scale and ideological network position, their focus on themes and narratives, and their discourse and language characteristics.
    
    Figure~\ref{fig:diagram} shows  our data measurement pipeline.
    Informed by the scant existing literature on tankies, we set out to answer these research questions by first identifying a set of tankie subreddits.
    We then measure over 1M posts from 50K authors in our dataset across a variety of axes, giving us a unique view of how tankies are positioned within the larger left-wing community.
    First, we perform a set of quantitative analyses that reveal the relationship between tankies, other far-left communities, leftists, feminists, and capitalists.
    We then measure user migrations between tankies and other far-left communities.
    Further, we look at how tankies compare to the rest of the far-left with respect to their vocabulary and the topics they discuss.
    We also analyze who they discuss, the domains they share, and the toxicity of their discussions to expand our findings.
    By using comprehensive measurement methodologies, we  provide empirical evidence and insights into the behaviors and influences of left-wing extremist communities on social media.

    \descr{Contributions.}
    In summary, this paper makes several contributions.
    First, we provide the first large-scale measurements of the far-left on social media and introduce a quantitative framework for discovering and measuring far-left extremists.
    We identify and quantify the tankie activity on Reddit. 
    We find that tankies had more active users than any other far-left subreddits, their user base has been steadily increasing over time, and that there is a drop in monthly active users in every far-left subreddit except tankies after the Russian invasion of Ukraine (\textbf{RQ1}).
    Additionally, we measure user migration patterns from other far-left communities to tankies, which coincided with tankies becoming the most popular far-left community on Reddit prior to being banned (\textbf{RQ1}).
    We then characterize the tankie community on Reddit. To the best of our knowledge, this paper provides the first social media characterization of tankies in the literature across a multitude of disciplines.
    When looking at tankies' positioning within our network of ideological subreddits, we find they are indeed at the periphery of the far-left (\textbf{RQ2}), and should thus be considered left-wing extremists.
    We further show that tankies differ from the rest of the far-left in several striking ways.
    By developing a metric to measure topic prevalence, we find that tankies give less priority to core left-wing issues like healthcare and housing, instead prioritizing state-level political discussions (\textbf{RQ3}).
    Finally, we measure linguistic and behavioral differences, finding that tankies show consistent vocabulary misalignments when compared to other far-left communities, and share more links from revisionist news sources (\textbf{RQ3}).
    We also find that tankies exhibit high toxicity and use pejorative language, in line with prior work on other online extremist communities~\cite{hine2017kek,zannettou2018origins, grover2019detecting} (\textbf{RQ3}).
     By situating tankies within the broader far-left, our study shows how authoritarian and geopolitically aligned extremists emerges and differentiates itself from other left-wing communities, despite sharing many ideological reference points with them.

    \section{Background \& Related Work}
    \label{sec:background}

    The rise of online social media platforms has had a substantial impact on the way individuals consume and produce political content. 
    In particular, the ability for individuals to self-select into ideologically-homogeneous online communities has led to the proliferation of extremist groups on the Internet. 
    Although research on right-wing extremist groups on social media is extensive, there has been relatively less focus on left-wing extremist groups. 
    In this paper, we aim to fill this gap by examining the online behavior of a left-wing extremist community known as ``tankies.''
    
    \descr{What is a ``tankie?''}
    \emph{Tankie} was originally a pejorative term referring to communists who supported the USSR's invasion of Hungary in 1956 and Czechoslovakia in 1968~\cite{robinson2011new,petterson2020apostles}.
    Over the years, the context of the usage of tankie evolved.
    For example, it has been used to express derision towards pro-Soviet hardliners~\cite{glastonbury1998children}, to describe communists who support China's policies~\cite{lanza2021rose} (e.g., supporters of China's actions on Uyghurs~\cite{roberts2020war} and the Hong Kong protests~\cite{andersenchan2021}), as well as young, online Stalinists in general~\cite{greenwalt2022get}.
    
    Thus, tankie is now used to describe much more than the set of communists who supported specific events from the Soviet era.
    The term tankie now covers communists who support ``Actually Existing Socialist'' (AES) countries; especially those with a Stalinist or authoritarian leaning.
    Although there is not really a concrete definition, recent work by Petterson~\cite{petterson2020apostles} provides a succinct description of tankies:
    \begin{quote}
        ``Tankies regard past and current socialist systems as legitimate attempts at creating communism, and thus have not distanced themselves from Stalin, China etc.''
      \end{quote}

    \descr{Imbalance in Research on Online Extremism.}
    While research on online extremism has been increasing, there has been a particular focus on right-wing extremist communities, e.g., those affiliated with white supremacy or far-right politics~\cite{grover2019detecting, squire2021monetizing}.
    These communities can be found on various social media platforms, including forums, blogs, and social networking sites.
    Social media platforms like Reddit, 4chan, Gab, Parler, and others have been found to harbor far-right communities, providing a gathering place for like-minded individuals, recruiting and further radicalize others, and even organize violent, real-world activity~\cite{zannettou2017web,hine2017kek,guhl2020safe,zannettou2018origins}.
    Previous studies have demonstrated that domestic extremist groups in the United States, particularly those on the right-wing, tend to form networked clusters of websites that share similar ideologies.
    These clusters often contain prominent sites that connect different clusters, providing access and control to resources for members, sympathizers, and interested organizations~\cite{zhou2005us,reid2007internet}.
    Another study that analyzes the posting behaviors in one of these websites finds that the messages are assertive in tone and contain a variety of emotionally charged language, including words like ``bomb,'' ``kill,'' ``evil,'' and ``threat,'' that are directed towards specific adversary groups, and often advocates for violence against Jews~\cite{scrivens2021exploring}. 
    
    In contrast, research on online extremist left communities is relatively limited~\cite{didrivers}. 
    Studies have found that far-left extremist offenses share commonalities with common criminal acts like vandalism, theft, and arson~\cite{sproles2019creativity}, and a negative correlation between the timing of their cyberattacks and physical violence~\cite{holt2021examining}. 
    While there have been studies that analyze online far-left communities, they generally involve non-extremist far-left communities (e.g., supporters of European socialist parties), and are often limited to a specific country. 
    E.g., a study of Greek far-left and far-right tweets did not find significant differences in network structure and sentiment during three different politically intense time periods~\cite{agathangelou2017understanding}.
    Another study found that both online Australian far-right and far-left communities become more active in response to major global events~\cite{davey2022far}.
    Additionally, a prior study reported that East Germany consistently ranked slightly higher than the West in terms of various extreme left-wing attitudes over time, despite the fact that absolute numbers have recently declined~\cite{jungkunz2019towards}.
    Although there are other data-driven studies that capture far-left groups~\cite{zihiri2022qanon,klein2019twitter}, their primary focus is on the far-right.
    These studies use far-left groups as a comparative backdrop, not examining their intrinsic characteristics.
    To our knowledge, no large-scale data-driven studies have analyzed left-wing extremists on social media. 
    Therefore, it is crucial to study left-wing extremism online to fill the knowledge gap and better understand all forms of extremism, which is essential for effective countermeasures.
    \begin{figure}[t]
      \centering
      \includegraphics[width=0.8\columnwidth]{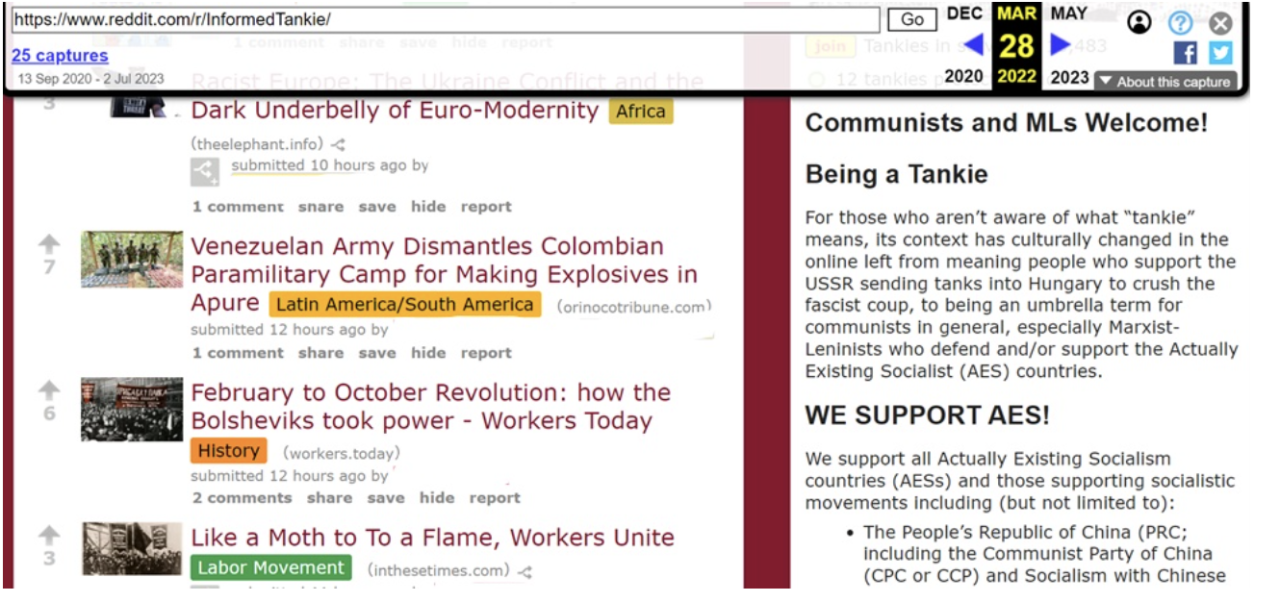}
      \caption{
              Screenshot of r/InformedTankie.}
      \label{fig:informed_tankie}
  \end{figure}
    
    \section{Measuring Tankie Activity}
    \label{sec:data}
    
    Left-wing extremism is an understudied area in comparison to right-wing extremism. 
    One of the reasons for this is that identifying online communities associated with left-wing extremist groups is challenging in and of itself.
    To address this gap, we conduct a systematic measurement of tankie activity on Reddit. 
    Our first step is to identify and measure a group of subreddits that are part of the overall tankie community (RQ1), which represents a segment of the broader left-wing extremist landscape.
    Then, to explore how tankies are positioned within the overall political discourse on Reddit, we collect and analyze data from subreddits representing mainstream ideologies.
    This allows us to measure the relative positioning of tankies within the wider ideological spectrum.
    Lastly, we measure the community growth of tankies by analyzing the monthly active user counts and migration patterns within the larger far-left community to quantify the size and activity level of its user base (RQ1).
    
    \descr{Why Reddit?}
    Reddit, with over 1.5 billion monthly visits as of March 2022~\cite{statista2022}, is among the world's leading social platforms.
    It comprises over 130,000 ``subreddits,'' each with its own rules and moderated by volunteers. 
    Users engage by posting links, commenting, and voting.
    Each subreddit can feature a banner, often conveying its theme or rules. 
    Reddit's variety of political subreddits provides insights into tankies' positioning in the political landscape. 
    Given its user-driven communities, Reddit has been studied for extremist groups and moderation strategies~\cite{ribeiro2020evolution,mamie2021anti,aliapoulios2021gospel,horta2021platform,chandrasekharan2022quarantined}, making it ideal for analyzing the tankie community.
    
    \begin{table}[t]
      \begin{scriptsize}
          \begin{center}
          
          \begin{tabular}{lrrrrr}
              \toprule
              \multicolumn{1}{c}{Subreddit} & \multicolumn{1}{c}{\# Posts} & \multicolumn{1}{c}{\# Authors}  & \multicolumn{1}{c}{\begin{tabular}[c]{@{}c@{}}Min - Max\\ Date\end{tabular}}      \\ 
              \midrule
              r/GenZedong                   & 1,273,277                   & 50,153                         & 07/2019 - 03/2022                                                                      \\
              r/sendinthetanks              & 42,813                      & 5,603                          & 03/2020 - 03/2022                                                                       \\
              r/InformedTankie              & 29,242                      & 3,697                          & 03/2020 - 03/2022                                                                          \\
              r/GenZhou                     & 28,606                      & 3,620                          & 12/2020 - 03/2022                                                                     &  \\
              r/asktankies                  & 6,678                       & 937                            & 01/2021 - 03/2022                                                                       \\
              r/BunkerTube                  & 12                          & 5                              & 11/2020 - 01/2022                                                                      \\ 
          \bottomrule    
          \end{tabular}
          
          \end{center}
      \end{scriptsize}
      \caption{General information of the tankie subreddits from identified reference network. }
    \label{tab:tankie_infos}
      \end{table}

    \descr{Identifying tankie subreddits.}
    To our knowledge, this is the first large-scale measurement study to examine tankies.
    Although previous research~\cite{grover2019detecting,waller2021quantifying} \emph{does} mention subreddits r/tankie and r/tankies, both have relatively few subscribers (42 and 213 at the time of this writing, respectively).
    We thus create our own tankie subreddit list.
    
    We begin by searching for the terms ``tankie'' and ``tankies'' on Reddit, discovering \emph{r/InformedTankie} with 13.8K members.
    Via a qualitative analysis, it is clear that r/InformedTankie is a central hub for tankies, offering resources and insights into their ideology (screenshot can be seen in Figure~\ref{fig:informed_tankie}).
    This includes providing a wiki, promoting a Discord server for tankie users, and having a dedicated section for supporting Actually Existing Socialist (AES) countries.
    For example, r/InformedTankie's banner defines ``tankie'' in line with Petterson's description:
    \begin{quote}
        For those who aren't aware of what ``tankie'' means, its context has culturally changed in the online left from meaning people who support the USSR sending tanks into Hungary to crush the fascist coup, to being an umbrella term for communists in general, especially Marxist-Leninists who defend and/or support the Actually Existing Socialist (AES) countries.
      \end{quote}
    The banner also includes a list of eight ``Allied Subreddits.''
    Via manual inspection, we find that four of them (r/Genzedong, r/sendinthetanks, r/asktankies, and r/BunkerTube) express support for so-called AES countries and identify as communist (specifically, ``Marxist'' or ``Marxist-Leninist'') in their banners.
    The remaining four subreddits focus on broader Marxist/Marxist-Leninist views rather than being specifically tankie related.    
    
    While our findings (including those later in this section) suggest that r/InformedTankie serves as a central hub within the tankie community, it is important to acknowledge the possibility of other tankie subreddits existing outside this network. 
    These unlinked subreddits may also exhibit tankie characteristics, even though they are not directly associated with r/InformedTankie.
    To discover additional tankie subreddits, we follow a manual snowball sampling strategy.
    Starting from r/\hspace{0pt}InformedTankie, we recursively look for additional subreddits linked to in banners that fit with Petterson's description and the definition from r/InformedTankie.  The search stops when we no longer encounter any additional tankie subreddits within the next three hops.
    This network includes one additional tankie subreddit, r/GenZhou.

    Recognizing that our initial approach might have missed other relevant tankie subreddits, we expanded our efforts to uncover additional communities that may exist between the six identified tankie subreddits and the broader communist community on Reddit.
    To explore this, we build a comprehensive ``reference network.'' 
    We build a simple crawler that recursively collects the names of any subreddits listed in the banners of the subreddits we have collected so far, stopping when we discover no new subreddits.
    This process results in a large network of 21,161 subreddits and 114,962 edges, where edges represent links between subreddits in banners.

    Since tankies position themselves within the larger ideology of communism, we create a subgraph of our reference network composed of the shortest paths between our six tankie subreddits and r/communism, pinpointing the most directly connected subreddits. 
    The resulting subgraph has 18 nodes (i.e., subreddits) and 19 edges, but does not contain any additional tankie subreddits.
    Table~\ref{tab:tankie_infos} shows general information of the tankie subreddits we identify.
    
    To better understand the prominence and connectivity of the tankie subreddits in our reference network, we compute three different network measures, in-degree, out-degree, and PageRank (explained in Appendix~\ref{sec:appendix_data}).
    This allows us to gain deeper insight into the online behavior and interactions of tankies.
    Notably, r/GenZedong is the most popular tankie subreddit, with nearly 1.3M posts and 50K authors.
    Despite having zero out-degree score, r/GenZedong has the highest in-degree and ranks second in PageRank, highlighting its dominant presence.
    Due to spreading misinformation, r/GenZedong was quarantined by Reddit on March 23, 2022~\cite{redditgenzedong}.
    Furthermore, r/InformedTankie references 4 out of the 5 identified tankie subreddits (excluding itself), the highest among the identified tankie subreddits (refer to Table~\ref{tab:tankies_network_subs} and Table~\ref{tab:network_subs_definition} in Appendix~\ref{sec:appendix_data}).
    Further details about the tankie subreddits and the subreddits linking to them can be found in Appendix~\ref{sec:appendix_data}.

    \descr{Identifying ideology subreddits.}
    To understand tankies' ideological stance, we initially need to identify prominent ideological subreddits, specifically those leaning left. 
    To accomplish this, we use a list~\cite{redditlist} previously applied for such identifications~\cite{shen2021sounds}.
    Subsequently, we carry out a comparative analysis between communities from this list that demonstrate similarities to the tankie ideology.
    This list has 43 subreddits representing a range of ideologies, from r/Conservative to r/Anarchism, including numerous subvariants, e.g., r/\hspace{0pt}AnarchoPacifism.
    After manual examination, we remove one non-English subreddit (r/piratenpartei) and those that are defunct or private.
    Further manual examination shows r/Communist, r/Anarchist, and r/feminisms are not the largest subreddits for their respective ideologies.
    So, we incorporate the largest subreddits explicitly dedicated to them (r/communism, r/\hspace{0pt}Anarchism, and r/Feminism).

    \descr{Post collection.}
    We first collect all posts from the six tankie subreddits we identify earlier in this section using the Pushshift API~~\cite{baumgartner2020pushshift}.
    As mentioned previously, r/GenZedong was quarantined due to spreading misinformation on March 23, 2022.
    Hence, our dataset spans from July 15, 2019 (marking the earliest tankie subreddit post) up to the quarantine date.
    Cumulatively, our dataset contains 22,481,036 posts, with 1,380,628 from tankies and the remaining 21,100,408 from other ideological subreddits.
    For comparison, a far-right subreddit, r/The\_Donald, that was banned from reddit for violating Reddit's rules against harassment, hate speech, and content manipulation~\cite{Allyn_2020}, had approximately 3 million posts between October 29, 2019 and February 26, 2020~\cite{horta2021platform}.
    Figure~\ref{fig:total_number_of_posts} shows the distribution of these posts across subreddits.

    \begin{figure}[t]
        \centering
        \includegraphics[width=1\columnwidth]{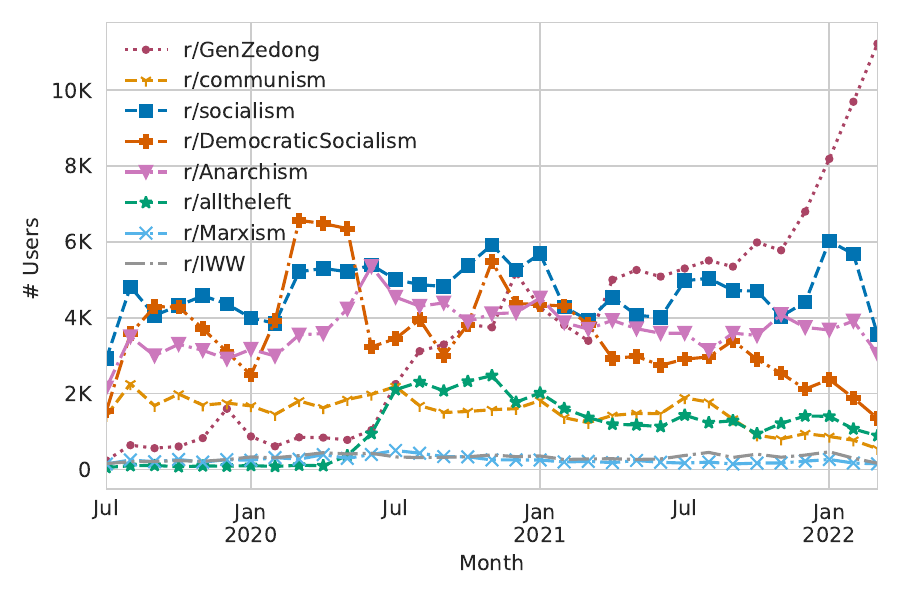}
        \caption{
                Monthly active user counts of the largest tankie subreddit, r/GenZedong, compared to non-tankie far-left communities. r/GenZedong outgrows the monthly active user counts of any other far-left community starting from April 2021.}
        \label{fig:monthly_user_counts}
    \end{figure}
    
    \begin{figure*}[ht]
      \centering
      \includegraphics[width=1\textwidth]{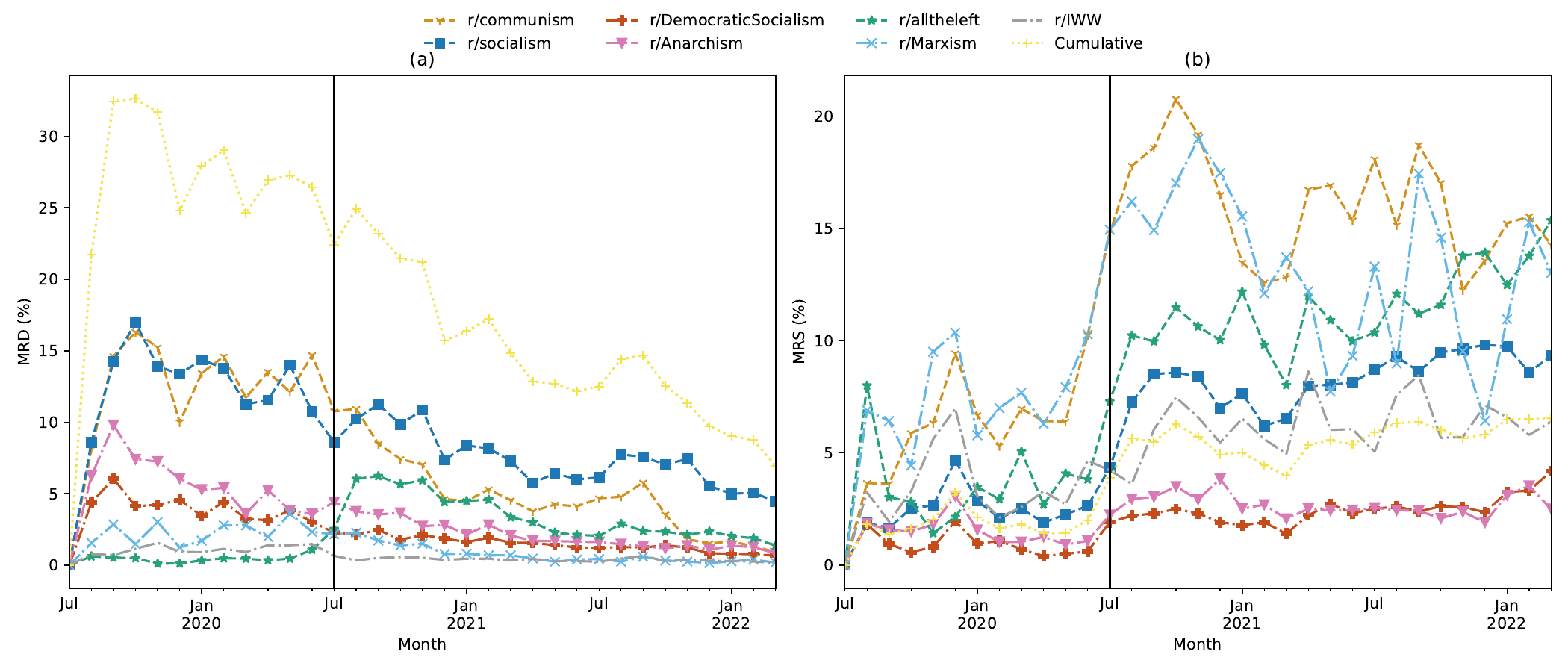}
      \caption{Monthly migrations from far-left subreddits relative to tankie subreddits (a) and monthly migrations to tankie subreddits relative to far-left subreddits (b). The black line shows the month tankies surpass the monthly user count of r/communism. The figure illustrates the rapid influx of members to the tankies from other far-left communities once it surpassed the user base of r/communism. }
      \label{fig:monthly_migrations}
    \end{figure*}

    \descr{Community Growth.}
    Despite being an old community, tankies have historically exerted minimal tangible influence within the broader context.
    To understand how things have changed, we first measure the growth and prevalence of the tankies community on Reddit compared to other far-left communities (RQ1) by computing the monthly active user counts (MAU) for each far-left community in our dataset.
    Specifically, for each far-left subreddit and month, we compute the monthly active user counts (MAU), defined as the number of distinct accounts that post at least once in that subreddit during that month. 
    We then compare the MAU trajectory of tankies to that of other far-left subreddits, allowing us to assess whethere tankies remain a marginal niche or have become one of the main hubs of far-left.
    In the following section (Section~\ref{sec:user_base}), we detail our approach to identifying these far-left communities. 
    Subsequently, we measure migration patterns from other far-left communities to tankies to better understand tankies' growth within far-left ideologies.
    
    Figure~\ref{fig:monthly_user_counts} shows the MAUs of the largest tankie subreddit, r/GenZedong, compared to non-tankie subreddits.
    Additional data, including other tankie subreddits and cumulative tankie user counts, are presented in Figure~\ref{fig:appendix_monthly_user_counts} in Appendix~\ref{sec:appendix_data}.
    Starting in July 2020, the tankies community saw a large influx of active users, doubling in about six months and growing larger than the user base of r/communism.
    As of April 2021, tankies had more active users than any other far-left subreddits.
    MAU for tankies doubles again between November 2021 and March 2022, reaching 11,872 active users.
    r/socialism has the second highest MAU of far-left subreddits after tankies with 3,559 (30\% of tankies for that month).
    We also see a decline in monthly active users in every far-left subreddit \emph{except} tankies starting in February 2022, which is the month Russia invaded Ukraine~\cite{johnmurphy2022}.

    To understand the significance of the tankies community's growth, we run a Mann-Kendall Test~\cite{mann1945nonparametric,kendall1948rank}, a non-parametric test for analyzing monotonic trends in time series data, on our MAU time series.
    For $\alpha = 0.01$, we find tankies are the only community with a statistically significant increase in MAU ($z > 0$ and $|z|>z(1-\alpha/2)$, $p < 0.01$).
    We also observe a strong positive correlation between tankies' MAU and time ( $\rho = 0.97$, $p < 0.01$) when applying Spearman's Correlation, indicating that the tankies user base has been steadily increasing over time.
    These results raise an important question: are these completely new users getting involved in far-left ideologies, or is the user base of existing far-left subreddits migrating to tankies over time?

    Using the approach outlined by Ribeiro et al.~\cite{ribeiro2020evolution}, which measures the movement of users across different Manosphere (online communities centered around men's rights and masculinist ideologies~\cite{lilly2016world, ribeiro2020evolution}) subreddits, we measure the extent of migration from a given source community to a destination community via two metrics:

    \descr{Migration Relative to Destination (MRD)}, which measures the degree to which migrations \emph{into} a destination community can be attributed to a given source community:

    \[MRD = \frac{S_{t-1} \cap D_t}{D_t}\]

    \descr{Migration Relative to Source (MRS)}, which measures the degree to which migrations \emph{from} a source community can be attributed to a given destination community:

    \[MRS = \frac{S_{t-1} \cap D_t}{S_{t-1}}\]

    where $D_t$ is the number of authors posting in a given \emph{destination} community in a given month $t$ and $S_{t-1}$ is the number of authors that posted in a given \emph{source} community in the previous month.

    In this analysis, we examine two distinct time periods: July 2019 (when r/GenZedong was founded) to June 2020 and July 2020 (when the tankies community overtook r/communism in terms of MAU) to March 2022.  
    When we examine the migratory patterns between tankies and other far-left communities in Figure~\ref{fig:monthly_migrations}, we find that the popularity of tankies increased among other far-left communities (except r/\hspace{0pt}Anarchism), even though their proportional contribution to the user base of tankies decreased over time.
    This suggests that the initial tankies community started with a relatively small number of members from established far-left communities, and then rapidly siphoned off members from these other far-left communities once it exceeded the user base of r/communism (for in-depth exploration, see Appendix~\ref{sec:appendix_data}.)

    \descr{Takeaways.}
    Tankies on Reddit experienced a notable growth in monthly active users (MAU) since July 2020, surpassing other far-left subreddits in user activity from April 2021 onwards.
    This indicates that, before facing restrictions on Reddit, tankies became more popular than the other far-left communities we examined (RQ1).
    Furthermore, while other far-left subreddits experienced a decline in MAU starting in February 2022, the month Russia invaded Ukraine, tankies continued their growth trajectory.
    This divergence suggests that tankies may have been uniquely impacted by the geopolitical events of that period.
    One possible explanation for this growth could be the alignment of tankie ideologies with pro-Russian narratives (as we show later in Section~\ref{sec:content_analysis}), which might have attracted users sympathetic to these views during the conflict. 
    Alternatively, the growth could be influenced by coordinated efforts from external actors seeking to amplify certain ideological perspectives, though further investigation would be needed to substantiate such claims. 
    Additionally, their migratory patterns indicate that tankies became less reliant on established far-left communities once they reached a certain level of user base.
    Taken together, these growth and migration patterns show that tankies evolved from a relatively marginal subculture into a rapidly expanding hub within the far-left.
    This justifies our subsequent focus on their position and discourse, since the distinctive behaviors we document in later sections concern a community whose audience is actively expanding rather than a static fringe.

    \begin{figure*}[t]
      \centering
      \begin{subfigure}[b]{0.5\textwidth}
          \centering
          \includegraphics[width=\linewidth]{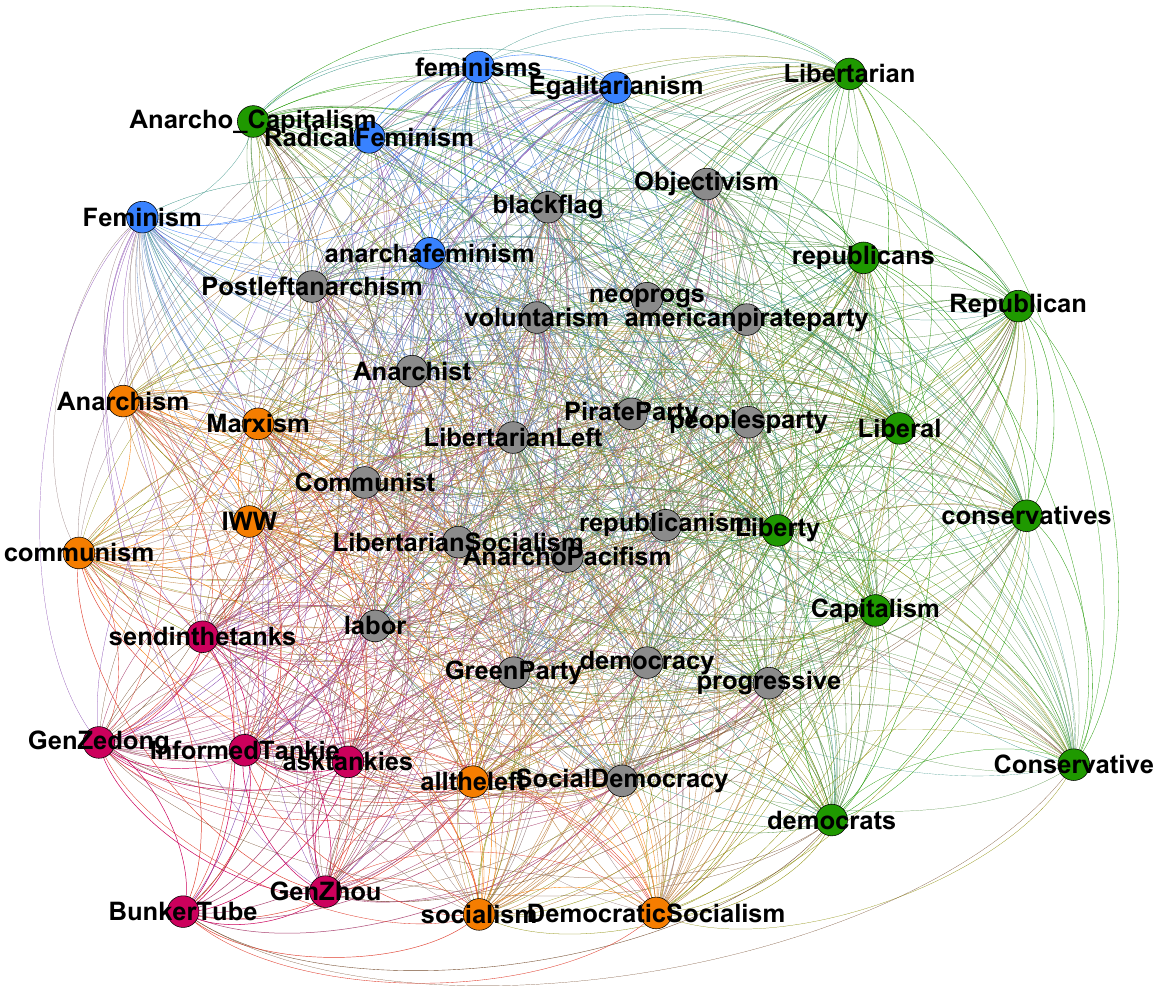}
          \caption{Ideological subreddit network.}
          \label{fig:atlas2_all_colored}
      \end{subfigure}
      \begin{subfigure}[b]{0.5\textwidth}
          \centering
          \includegraphics[width=\linewidth]{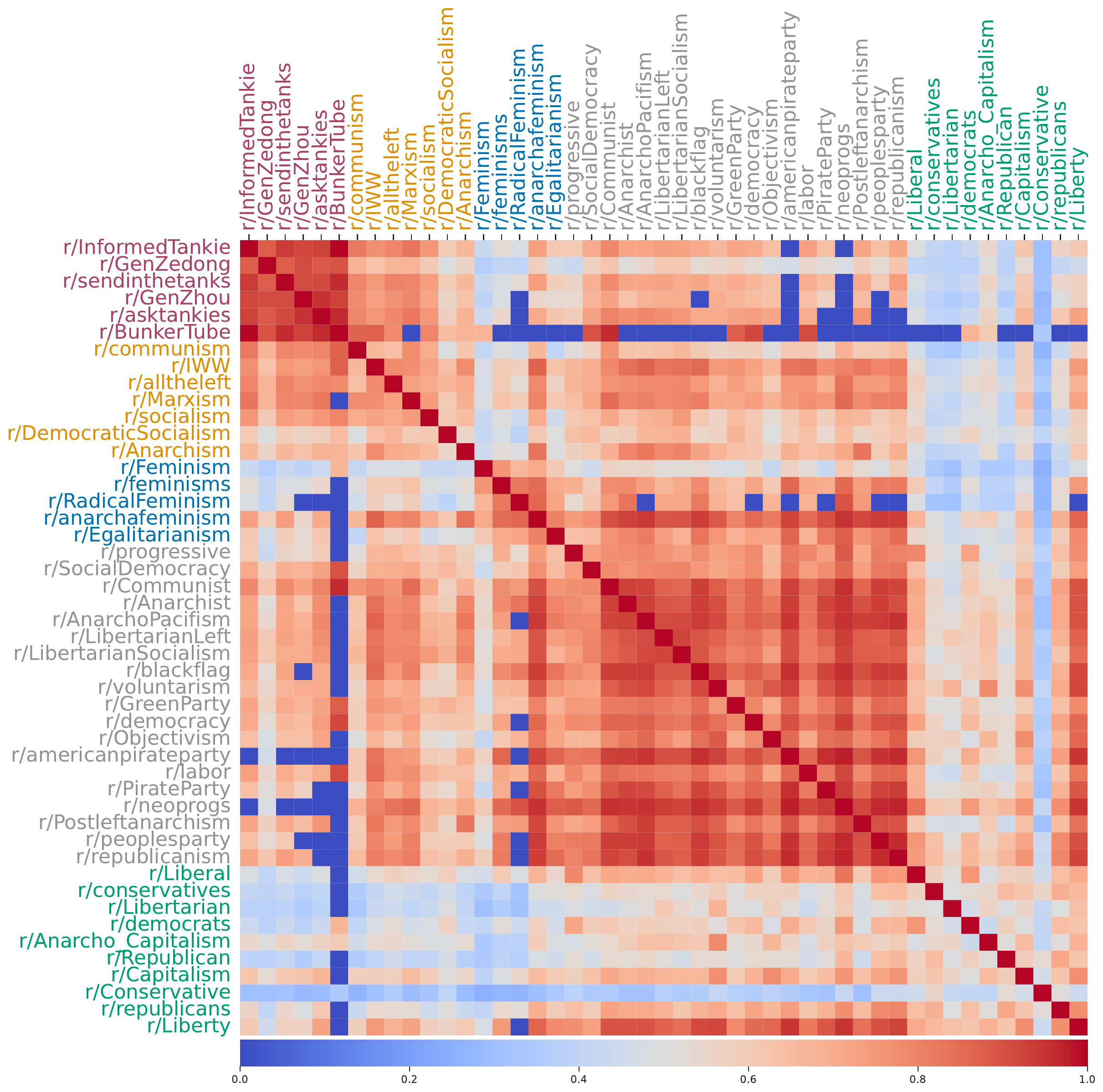}
          \caption{Standardized user overlap of ideological subreddits.}
          \label{fig:normalized_heatmap}
      \end{subfigure}
      
         \caption{The relationship between user bases of ideological subreddits. Far-Left subreddits are colored in red (with non-tankies colored light red), Leftists are colored in gray, Capitalists are colored in green, and Feminists are colored in blue. A stronger red hue indicates a higher similarity between the user bases of two subreddits.}
         \label{fig:bonacich_figures}
    \end{figure*}

    \section{Ideological Position of Tankies}
    \label{sec:user_base}
    In this section, we measure the prominence of tankies within the far-left ideology spectrum and their engagement with other far-left communities (RQ2).
    To achieve this, we construct a graph of subreddits based on user posting activity, detect communities to identify clusters related to tankie subreddits, and measure user base similarities between other subreddits in our dataset.

    \descr{Graph Construction \& Community Detection}
    Our first step is to detect communities of similar ideological subreddits.
    To do this, we follow a methodology similar to that of~\cite{Squire2019Network}.
    First, we construct a graph where each node represents a subreddit and the edges between the nodes represent the number of users they have in common. 
    These edges are weighted using Bonacich's method~\cite{bonacich1972technique}, which normalizes co-membership based on the full $2 \times 2$ contingency table of shared and non-shared members across each pair of subreddits.
    This controls for differences in community size and produces a normalized coefficient in $[0,1]$ that reflects the strength of membership overlap between two communities. 
    This methodology allows us to distinguish meaningful structural ties from overlaps that arise simply because some subreddits are much larger than others, unveiling structural patterns not evident in unprocessed overlap data.

    Next, we apply the Louvain community detection algorithm~\cite{blondel2008fast} to identify clusters of subreddits within the network. 
    This algorithm, works by optimizing a modularity measure, assessing how well the network is divided into distinct communities.
    It is widely used and validated for community detection in networks, particularly in social networks~\cite{ghosh2018distributed, hric2014community}.
    We expect that if our subreddits do indeed represent different political ideologies, the structure of this graph will naturally lead to empirical clusters that align with the theoretical literature.
    To check if these clusters can be further divided, we continuously apply the Louvain method on each cluster, adjusting the edges for every subcluster, until no new clusters emerge. 
    This allows us to examine the full spectrum of ideologies represented in our dataset.

    \descr{Results.}
    Figure~\ref{fig:atlas2_all_colored} plots the resulting graph, where nodes are colored according to the community they were placed in, and the graph is laid out in space such that subreddits that share fewer common authors after standardization are farther apart from each other~\cite{10.1371/journal.pone.0098679}.
    We next describe the clusters in more detail.
    
    \descr{Far-Left (Red Cluster).}
    Building on March's definition of the far-left~\cite{march2008contemporary}, we label this cluster as far-left, since 
    tankies are indeed clustered with subreddits representing well-known far-left ideologies, e.g., r/communism, r/socialism, r/Anarchism, r/Marxism, and r/IWW (Industrial Workers of the World, a syndicalist labor union~\cite{cole2017wobblies}), representing nearly 9,000 workers across North America~\cite{iww}.
    We also see r/alltheleft in this cluster, which has a far-left identity, similar to communist, socialist, and anarchist subreddits~\cite{massoc2021social}. This cluster does not split into more clusters when we recursively apply the Louvain algorithm.
    
    \descr{Other Ideologies.}
    While the far-left communities were neatly formed in the first round of clustering, other ideologies ended up in a large cluster together.
    We thus apply the Louvain algorithm recursively until subclusters do not split more, as previously described.
    In the end, we find three coherent clusters that fit into ideologies that have been well covered in the theoretical literature:
    \begin{itemize}
      \item \textbf{Feminists (Blue Cluster)}: This cluster includes feminist and feminist-leaning subreddits (r/feminism, r/feminisms, r/anarchafeminism, r/radicalfeminism, and r/egalitarianism).
      \item \textbf{Leftists (Gray Cluster)}: This cluster includes left-leaning ideologies that are closer to moderate/center-left on the ideology spectrum (e.g., r/socialdemocracy, r\slash greenparty, r\slash progressive), as well as libertarian-left subreddits (r/libertariansocialism and r/libertarianleft) and some anarcho/anarchist-leaning communities (e.g., r/anarchopacifism, r/voluntarism, and r/\hspace{0pt}republicanism).
      Notably, this cluster also contains several less popular far-left subreddits, e.g., r/Communist, that are often overlooked (see Figure~\ref{fig:total_number_of_posts} in the Appendix), highlighting their distinction from the more prominent far-left communities.
      \item \textbf{Capitalists (Green Cluster)}: This cluster includes liberal-leaning ideologies (e.g., r/liberal, r/democrats), anti-authoritarian capitalist ideologies (r/libertarian, r/anarchocapitalism), and conservative ideologies (e.g., r/conservative, r/republican) alongside r/capitalism.
      \end{itemize}
    
    When we look at how tankies and other ideology subreddits are laid out in Figure~\ref{fig:atlas2_all_colored}, we see some meaningful positions.
    For example, we can see that r/democrats is positioned between r/Capitalism, r/progressive, r/SocialDemocracy, r\slash DemocraticSocialism, and r/Conservative.
    This may seem odd at first glance, but considering that these ideologies represent current US politics, it makes sense that they are positioned close to each other.
    Additionally, even though one of the closest subreddits to r/conservatives is r/liberal, they are still relatively distant.
    Tankie subreddits are also positioned on the periphery of the far-left cluster, indicating their detachment from more moderate or mainstream far-left ideologies and highlighting their extremist stance.

    \begin{table*}[t]
      \centering
      \tiny
      \begin{tabularx}{\textwidth}{r r X r r | r r X r r}
          \toprule
          \multirow{1}{*}{\bf no.} & \multirow{1}{0.5cm}{\bf \#Posts} & \multirow{1}{3cm}{\bf Top 5 Words} & \multirow{1}{*}{\bf \%Posts} & \multirow{1}{*}{\bf RTPR} & \multirow{1}{*}{\bf no.} & \multirow{1}{0.5cm}{\bf \#Posts} & \multirow{1}{3cm}{\bf Top 5 Words} & \multirow{1}{*}{\bf \%Posts} & \multirow{1}{*}{\bf RTPR}  \\

          \midrule
          \cellcolor{lightteal}1                         & \cellcolor{lightteal}15,360            &  \cellcolor{lightteal}\textbf{genocide, xinjiang, camps, the, china}     & \cellcolor{lightteal}1.11  & \cellcolor{lightteal}1                      &26  & 5,265              & \textbf{lol, wtf, real, holy, wait}                 & 0.38 & 1                             \\
          2                         & 12,056            &  \textbf{communism, no, inshallah, is, the}         & 0.87  & 1                      &27  & 5,256              & ok, hope, wish, you, sorry                 & 0.38 & 3                             \\
          3                         & 10,377            &  socialism, socialist, is, the, of                  & 0.75  & 6                      &28  & 5,205              & fucking, bot, fuck, sometimes, beep        & 0.37 & 4                             \\
          \cellcolor{lightteal}4                         & \cellcolor{lightteal}10,320            &  \cellcolor{lightteal}\textbf{korea, north, kim, dprk, korean}           & \cellcolor{lightteal}0.74  & \cellcolor{lightteal}1                      & \cellcolor{lightteal}29  & \cellcolor{lightteal}5,110              & \cellcolor{lightteal}\textbf{china, nukes, us, war, the}                 & \cellcolor{lightteal}0.37 & \cellcolor{lightteal}1                             \\
          5                         & 9,806             &  \textbf{media, chen, falun, gong, news}            & 0.71  & 1                      &30  & 4,861              & joke, submission, guidelines, rsocialism, this & 0.35 & 4                             \\
          6                         & 9,018             &  comrade, thanks, you, comrades, thank              & 0.65  & 3                      &31  & 4,595              & cuba, cuban, castro, the, fidel            & 0.33 & 3                             \\
          7                         & 8,781             &  thank, thanks, you, nice, good                     & 0.63  & 4                      &32  & 4,422              & \textbf{lgbt, trans, gay, the, and}                 & 0.32 & 1                             \\
          8                         & 8,464             &  \textbf{fascism, fascist, fascists, the, is}       & 0.61  & 1                      &33  & 4,166              & vote, voting, caucus, party, green         & 0.30 & 6                             \\
          9                         & 7,834             &  ussr, soviet, the, art, of                         & 0.56  & 2                      &34  & 4,153              & \textbf{source, uvredditdownloader, context, picture, where} & 0.30 & 1                             \\
          10                        & 7,777             &  capitalism, yes, why, capitalist, yep              & 0.56  & 5                      &35  & 4,085              & cool, hell, yeah, beautiful, awesome       & 0.29 & 3                             \\
          11                        & 7,772             &  \textbf{video, link, amp32amp32, feedback, dmca}   & 0.56  & 1                      &36  & 4,040              & thanks, hate, bro, remindme, sharing       & 0.29 & 2                             \\
          12                        & 7,694             &  \textbf{lmao, mad, ah, ha, interesting}            & 0.55  & 1                      & \cellcolor{lightteal}37  & \cellcolor{lightteal}3,853              & \cellcolor{lightteal}\textbf{israel, jewish, jews, palestine, the}       & \cellcolor{lightteal}0.27 & \cellcolor{lightteal}1                            \\
          13                        & 6,965             & \textbf{o7, translation, translate, salute, comrade} & 0.50 & 1                      &38  & 3,717              & propaganda, ngl, western, is, its          & 0.26 & 3                             \\
          14                        & 6,594             & \textbf{fuck, what, oh, you, this}                  & 0.47  & 1                      &39  & 3,447              & que, de, la, spanish, spain                & 0.24 & 4                             \\
          15                        & 6,542             & she, her, aoc, shes, warren                         & 0.47  & 3                      &40  & 3,295              & biden, trump, joe, is, he                  & 0.23 & 3                             \\
          16                        & 6,391             & \textbf{image, the, blue, transcription, red}       & 0.46  & 1                      &41  & 3,120              & exactly, elaborate, whats, what, happened  & 0.22 & 2                             \\
          17                        & 6,220             & \textbf{stalin, he, was, the, stalins}              & 0.45  & 1                      & \cellcolor{lightorange}42  & \cellcolor{lightorange}2,924              & \cellcolor{lightorange}climate, change, the, nuclear, to          & \cellcolor{lightorange}0.21 & \cellcolor{lightorange}5                             \\
          18                        & 5,860             & \textbf{twitter, tweet, delete, this, account}      & 0.42  & 1                      &43  & 2,850              & religion, religious, christian, jesus, of  & 0.20 & 4                             \\
          19                        & 5,860             & \textbf{imperialism, imperialist, whataboutism, fascist, is} & 0.42 & 1              & \cellcolor{lightorange}44  & \cellcolor{lightorange}2,816              & \cellcolor{lightorange}healthcare, insurance, insulin, the, for   & \cellcolor{lightorange}0.20 & \cellcolor{lightorange}4                             \\
          \cellcolor{lightorange}20                        & \cellcolor{lightorange}5,815             & \cellcolor{lightorange}police, cops, acab, the, to                         & \cellcolor{lightorange}0.42  & \cellcolor{lightorange}5                      & \cellcolor{lightorange}45  & \cellcolor{lightorange}2,769              & \cellcolor{lightorange}landlords, rent, landlord, housing, to     & \cellcolor{lightorange}0.20 & \cellcolor{lightorange}6                             \\
          21                        & 5,775             & \textbf{flag, usavevideo, flags, the, prc}          & 0.41 & 1                       & \cellcolor{lightorange}46  & \cellcolor{lightorange}2,412              & \cellcolor{lightorange}gun, guns, the, to, and                    & \cellcolor{lightorange}0.17 & \cellcolor{lightorange}6                             \\
          22                        & 5,697             & banned, sub, ban, pcm, rcommunism                   & 0.41 & 2                       &47  & 2,320              & vegan, animals, meat, veganism, animal     & 0.16 & 5                            \\
          23                        & 5,675             & book, read, of, the, and                            & 0.41 & 5                       & \cellcolor{lightorange}48  & \cellcolor{lightorange}2,248              & \cellcolor{lightorange}iww, union, organizing, the, to            & \cellcolor{lightorange}0.16 & \cellcolor{lightorange}5                             \\
          \cellcolor{lightteal}24                        & \cellcolor{lightteal}5,455             & \cellcolor{lightteal}\textbf{ukraine, azov, russia, the, ukrainian}      & \cellcolor{lightteal}0.39 & \cellcolor{lightteal}1                       & \cellcolor{lightorange}49  & \cellcolor{lightorange}1,506              & \cellcolor{lightorange}tax, taxes, pay, fair, the                 & \cellcolor{lightorange}0.10 & \cellcolor{lightorange}7                             \\
          25                        & 5,362             & tankies, tankie, fash, holocaust, the               & 0.38 & 2                       &50  & 1,170              & amazon, bezos, stock, jeff, to             & 0.08 & 7                             \\
        
          \bottomrule    
      \end{tabularx}
  
      \caption{Topics discussed by tankies, ordered by the number of posts they have in each topic, along with the total number of posts and their proportions. We also report the Relative Topic Prevalence Rank (RTPR) compared to other far-left communities, where a higher rank (1 being the highest) indicates that tankies give more prevalence to the respective topic compared to other far-left communities.
      We find that tankies give more attention to state-level political events (highlighted in light teal), as evidenced by their higher RTPRs in these topics. Conversely, they give less attention to social issues (highlighted in light orange), indicated by their lower RTPRs in these topics. This indicates that tankies prioritize state-level political events over social issues compared to other far-left communities.}
    \label{tab:tankies_topics}
      \end{table*}

    To get a different perspective on the relatedness of different ideological communities, we plot a heatmap in Figure~\ref{fig:normalized_heatmap} showing the edge values for each pair of subreddits in our ideological graph. 
    From this heatmap, we can see that the subreddits in the far-left and leftist clusters are relatively well-connected compared to those in the capitalist community.
    Additionally, we observe that tankie subreddits are clearly much better connected with each other than other subreddits in their corresponding community.
    This is strong evidence that tankies are a \emph{distinct} community within the far-left.

    \descr{Takeaways.}
    Overall, we find that the user base of tankies is definitively positioned on the periphery of far-left ideologies (RQ2).
    Our analysis reveals that users within tankie subreddits display stronger intraconnectivity than other subreddits in the broader far-left community, underscoring their unique identity.
    Furthermore, tankie subreddits exhibit limited cross-engagement with other far-left subreddits, suggesting a degree of ideological isolation. 
    This insularity may reinforce distinct narratives in tankie discourse, deepening their ideological separation from other left-wing groups.

    \section{Measuring The Tankie Narrative}
    \label{sec:content_analysis}
    
    To measure the distinct narrative of tankies compared to other far-left communities (RQ3), we conduct a content analysis using a variety of techniques to compare them with other far-left subreddits.
    Initially, we perform topic analysis to discern \emph{what} tankies discuss.
    Next, we dissect their lexical choices by comprehensively analyzing their language misalignment and conceptual homomorphisms.
    To further validate and enrich our findings, we examine the named entities referenced by these communities, assess the tone of their discourse through toxicity analysis, and investigate the news outlets they frequently share.
    Detailed results and supplementary analyses are provided in Appendices~\ref{sec:ner_analysis}, \ref{sec:perspective}, and \ref{sec:domain_main}.

    \subsection{What do tankies talk about?}\label{sec:what_do_tankies_talk_about}
    
    We begin by determining \emph{what} tankies talk about via topic modeling.
    We use BERTopic~\cite{grootendorst2022bertopic}, a transformer-based model, to extract topics across documents. 
    Details about this technique as well as training details can be found in Appendix~\ref{sec:appendix_topic}.
    Due to the large number of clusters produced by the underlying technique of BERTopic, following~\cite{shen2022xing}, we hierarchically reduce clusters to a range of 20-80 (incrementing by 10 at each step) and select the model with the highest coherence score; 50 in our case.

    \begin{figure}[th!]
      \centering
      \includegraphics[width=1\columnwidth]{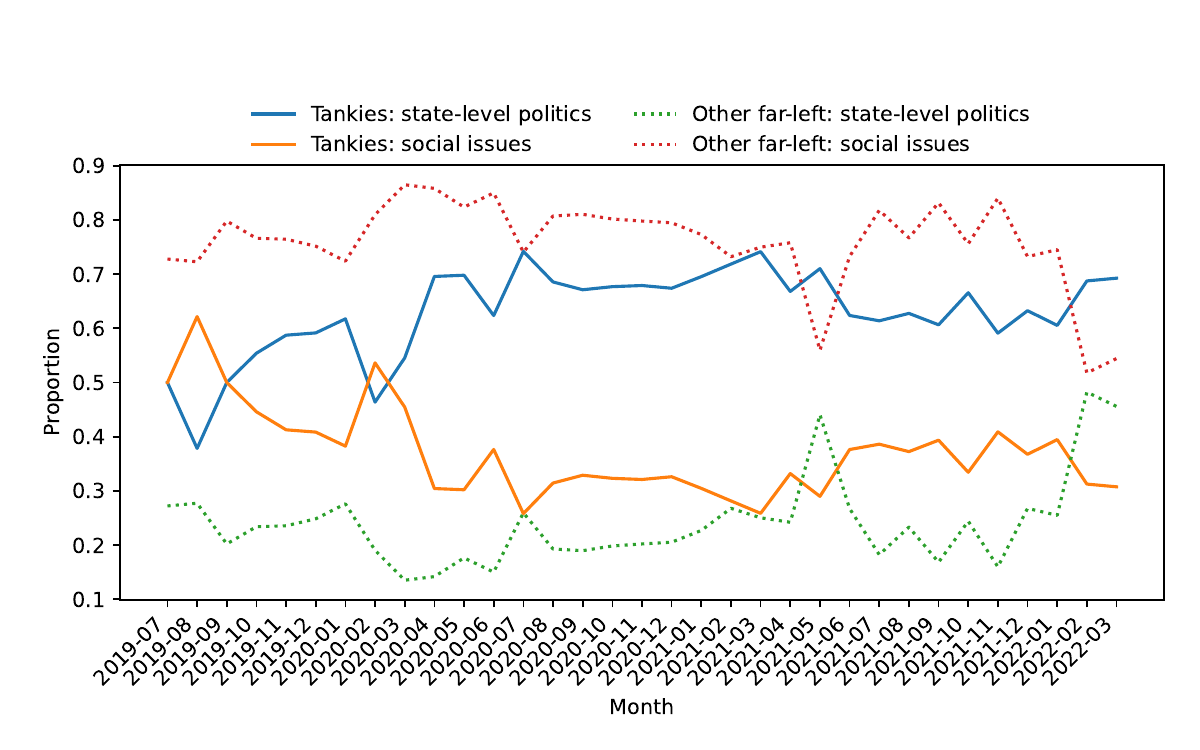}
      \caption{
              Monthly proportion of posts in highlighted topics assigned to state-level politics and core social issues for tankies and other far-left communities. Tankies consistently devote a larger share of these posts to state-level politics, while other far-left communities focus more on social issues. }
      \label{fig:dynamic_topic}
  \end{figure}
    
    To measure the prominence of tankie discussions relative to those in other far-left communities, we  introduce a metric called Relative Topic Prevalence Rank (RTPR).
    \[
    RTPR_{\text{topic}}(C) = \text{Rank}\left( \frac{P_{\text{topic}}(C)}{P_{\text{total}}(C)}, \, \mathcal{S} \right)
    \]
    where:
    \begin{itemize}
        \item $C$ denotes a specific community.
        \item $P_{\text{topic}}(C)$ represents the count of posts in community $C$ pertaining to the specified topic.
        \item $P_{\text{total}}(C)$ is the total post count in community $C$.
        \item $\mathcal{S}$ symbolizes the set of all such proportions across the considered communities, i.e., \newline  
        $\mathcal{S} = \left\{ \frac{P_{\text{topic}}(C')}{P_{\text{total}}(C')} \mid C' \in \text{Communities} \right\}$.
        \item The function $\text{Rank}(\cdot, \mathcal{S})$ assigns a comparative rank to each community based on the specified topic's proportional prevalence, juxtaposed against the collective dataset $\mathcal{S}$.
    \end{itemize}
    This metric ranks each community based on the proportion of their posts that pertain to a specific topic.
    A higher RTPR indicates that a given community focuses on a particular topic more than another community does.
    By using RTPR, we can compare and interpret the discussions of tankies within far-left communities, highlighting their unique focus areas and preferences. 
    This metric allows us to understand better how tankies differ from other far-left groups in their discourse and 
    ideological leanings.

    \descr{Results.}
    Table~\ref{tab:tankies_topics} presents the topics discussed by tankies ordered by the number of posts in each topic, as well as tankies' RTPR for each topic.
    We find that the most popular topic among tankies is the Uyghur (a Turkic ethnic group from Xinjiang) Genocide occurring in Xinjiang, China, accounting for 1.11\% of all tankies posts. 
    In addition, we see that topics related to ideologies (socialism, communism, fascism, and capitalism) and two AES states (North Korea and USSR) in the top 10.
    When we look at the RTPRs, we find that they generally focus more on recent, state-level political events compared to other far-left communities (see highlighted topics in Table~\ref{tab:tankies_topics}).
    As we can see from the table, tankies have RTPR = 1 for the topics related to Uyghur genocide (\#1), North Korea (\#4), Russian invasion of Ukraine (\#24), China-US conflicts (\#29), Israel-Palestine conflict (\#37), communism (\#2), fascism (\#8), Stalin (\#17).

    Conversely, tankies rank in the bottom half when it comes to focusing on social problems.
    We find tankies have RTPRs between 4-7 on topics related to police (\#20), climate change (\#42), healthcare (\#44), housing (\#45), guns (\#46), unions (\#48), and taxes (\#49).
    Additionally, our analysis reveals that tankies have RTPRs 2-3 on topics related to USSR (\#9), tankies (\#25), and Cuba (\#31).

    Figure~\ref{fig:dynamic_topic} plots the monthly proportion of posts in the highlighted topics for tankies and other far-left communities.
    We find that tankies consistently prioritize state-level politics, with these topics typically accounting for 60-70\% of the highlighted topic posts, and social issues making up the remaining 30-40\%.
    In contrast, other far-left communities show the opposite pattern, where between 70-80\% of the highlighted topic posts concern core social issues, while only about 20-30\% focus on state-level politics.
    Notably, we observe a visible increase in attention to state-level politcs in February-March 2022, coinciding with the Russian invasion of Ukraine, but the relative ordering between tankies and other far-left communities remains unchanged.

    These results suggest that \emph{tankies prioritize discussions of state-level political events over the core ideological issues and social problems that have long concerned the far-left community.
    }
    
    \descr{Takeaways.}
    Our findings reveal that the predominant topic of discussion among tankies centers on the Uyghur genocide, which happens in an AES country, China.
    Tankies prioritize discussions of state-level political events over social issues (RQ3), as indicated by their higher relative topic prevalence ranks (RTPRs) for topics like the Uyghur genocide, the Russian invasion of Ukraine, and the Israel-Palestine conflict. 
    This is further supported by their lower RTPRs for social issues like policing, climate change, healthcare, housing, gun control, unions, and taxes compared to other far-left communities.

    \subsection{What are their lexical differences?}
    \label{sec:misalignment}
    
    So far, we discover that tankies focus more on state-level political events and less on social issues and ideological principles than other far-left communities.
    In this section, we investigate the differences in language use between tankies and other far-left communities.
    Given the diverse beliefs represented in left-wing extremism, conventional tools often fall short in effectively capturing their complexity~\cite{jungkunz2019towards}.
    To capture this nuance, we leverage the concept of \emph{alignment}.
    This involves mapping tankies' word usage to other far-left groups using multidimensional embeddings, providing a deeper understanding of their unique characteristics.
    
    \descr{Aligning Multidimensional Word Embeddings.}
      Understanding the language usage in polarized communities is challenging due to their distinct vocabularies for identical concepts.
      E.g., Democrats and Republicans are political adversaries.
      Thus, a Democrat might say that ``Republicans are America's greatest threat'' (and vice versa), making direct comparisons of their language less useful when it comes to polarizing topics.
      To address this, KhudaBukhsh et al.~\cite{khudabukhsh2021we} formulate the problem as a form of translation between two groups using the same natural language.
      While not every ``translation'' is as clear-cut as the previous toy example, this technique has seen refinement~\cite{milbauer2021aligning} to capture more complex, \emph{multidimensional} aspects of ideological differences.
      As such, it emerges as an ideal tool for distinguishing disparities between tankies and other left-wing ideologies.

      In a nutshell, we align two multidimensional word embeddings by creating a \emph{bilingual} lexicon to anchor the alignment.
      We then use a translation function to project the embeddings from one community to another, and once both vocabularies live in the same vector space, we compute cosine similarity between corresponding word vectors to quantify how similarly each term is used across communities.
      We focus on two concepts:

       \descr{\textbf{Misalignment}} reflects variations in describing the same concept, revealing opposing perspectives. It occurs when a word from the vocabulary of the source community (tankies, in our case) maps to a different word in the vocabulary of the target community (another far-left community). 
      This helps us capture the use of words with polarized meanings between the two communities.
    
       \descr{\textbf{Conceptual Homomorphism}} captures unique terminology within a community that conveys shared ideas differently, often indicating specific ideological or cultural nuances. In essence, conceptual homomorphism is  misalignment, with the added condition that the term from the source community's vocabulary must be unique to that community.
      This helps us capture the unique words and slang from the source community.

    \descr{Preprocessing and Training.}
      Following the same method as~\cite{milbauer2021aligning}, we tokenize each comment for each far-left community corpus, remove formatting tokens and hyperlinks, make all characters lowercase, and common bigrams and trigrams into single tokens to catch the possible usage of organization names.
      To stabilize the word embeddings, we oversample sentences from non-tankie far-left subreddits until they reach the same size as tankies.
      We then train Word2Vec skip-gram models~\cite{mikolov2013distributed} for each far-left community using 100 dimensions and a maximum vocabulary of 30,000 words. 
      To account for oversampling, we adjust the threshold for the number of times a word must appear - which is set to 5 for tankies - by multiplying it with the size increase factor for other far-left communities.
      We choose the top 5,000 words from the shared vocabulary of each community pair as ``anchor words.'' 
      This is based on the findings of Milbauer et al.\cite{milbauer2021aligning}, who showed that this selection provides similar accuracy to using the entire shared vocabulary.
      To translate the embeddings between the communities, we employ the MultiCCA, as it gives the best accuracy.
    
    \descr{Discovering Misalignments \& Conceptual Homomorphisms.}
    We look for misaligned and conceptually homomorphic word pairs (with tankies as the source community) that have a cosine similarity greater than 0.5 (27\% of all cosine similarities).
    This threshold has been previously adopted in studies using Word2Vec to achieve more reliable outcomes, demonstrating its effectiveness in various research contexts~\cite{xue2021matching,ge2017improving,nguyen2020content}.
    The CDF of cosine similarities of the misaligned and conceptually homomorphic pairs between tankies and r/communism, r/socialism, r/DemocraticSocialism, and r/Anarchism can be seen in Figure~\ref{fig:cosine_similarity_cdfs} in Appendix~\ref{sec:appendix_misalignment}.

    \begin{table*}[t]
      \begin{adjustbox}{width=\textwidth}
        \begin{tabular}{llr|lr|lr|lr}
            \toprule
                          &  \multicolumn{2}{c|}{r/communism}                      & \multicolumn{2}{c|}{r/socialism}                             & \multicolumn{2}{c|}{r/DemocraticSocialism}                           & \multicolumn{2}{c}{r/Anarchism}                                                \\
            \midrule
                          & \cellcolor{pink}{CPC - CCP}                      & \cellcolor{pink}.82 & \cellcolor{pink}{CPC - CCP}                        & \cellcolor{pink}.80    & \cellcolor{green}{Zionists - Jews}           & \cellcolor{green}.68           & \cellcolor{lightgray}{Ultras - Tankies}          & \cellcolor{lightgray}.78 \\
                          & \cellcolor{pink}{Rioters - Protesters}           & \cellcolor{pink}.74 & \cellcolor{pink}{Rioters - Protesters}             & \cellcolor{pink}.79    & \cellcolor{lightgray}{Succdems - Liberals}   & \cellcolor{lightgray}.66           & \cellcolor{lightgray}{Anarkiddies - Tankies}     & \cellcolor{lightgray}.72 \\
                          & \cellcolor{lightgray}{Anarkiddies - Anarchists}  & \cellcolor{lightgray}.65 & \cellcolor{lime}{Trot -Trotskyists}                & \cellcolor{lime}.76    & \cellcolor{pink}{Rioters - Protesters}       & \cellcolor{pink}.62           & \cellcolor{pink}{Rioters - Protestors}           & \cellcolor{pink}.72 \\
                          & \cellcolor{lime}{Trots - Trotskyists}            & \cellcolor{lime}.65 & \cellcolor{cyan}{LPR - Donetsk}                    & \cellcolor{cyan}.73    & \cellcolor{green}{Zionist - Israel}          & \cellcolor{green}.59           & \cellcolor{cyan}{LPR - Donbass}                  & \cellcolor{cyan}.69 \\
             Misalignment & \cellcolor{lime}{Destalinization - Khrushchev}   & \cellcolor{lime}.64 & \cellcolor{lime}{Khrushchev - Stalin}              & \cellcolor{lime}.72    & {Bidens - Trumps}                           & .59           & \cellcolor{lightgray}{Radlibs - Tankies}         & \cellcolor{lightgray}.69 \\
                          & \cellcolor{lime}{De-Stalinization - Khrushchev}  & \cellcolor{lime}.62 & \cellcolor{lightgray}{Succdems - Socdems}          & \cellcolor{lightgray}.72    & \cellcolor{green}{Zionism - Israel}          & \cellcolor{green}.56           & \cellcolor{pink}{CPC - CCP}                      & \cellcolor{pink}.66 \\
                          & \cellcolor{green}{Zionist - Israeli}             & \cellcolor{green}.60 & \cellcolor{cyan}{DPR - Donetsk}                    & \cellcolor{cyan}.66    & \cellcolor{orange}{RT - NYT}                 &  \cellcolor{orange}.53           & \cellcolor{lime}{Trotskysts - Stalinists}        & \cellcolor{lime}.64 \\        
                          & \cellcolor{cyan}{LPR - Donbass}                  & \cellcolor{cyan}.54 & \cellcolor{lightgray}{Anarkiddies - Anarchists}    & \cellcolor{lightgray}.64    & \cellcolor{pink}{Tiananmen Square - Riots}   & \cellcolor{pink}.52           & \cellcolor{green}{Zionist - Israel}              & \cellcolor{green}.62 \\       
                          & \cellcolor{cyan}{DPR - Donbass}                  & \cellcolor{cyan}.52 & \cellcolor{pink}{Reeducation Camps - Uighurs}      & \cellcolor{pink}.60    & {Popularity Contest - Elections}            & .51           & \cellcolor{cyan}{DPR - Donbass}                  & \cellcolor{cyan}.61 \\
                          & \cellcolor{pink}{Boarding Schools - Uyghurs}     & \cellcolor{pink}.52 & \cellcolor{pink}{Internment Camps - Camps}         & \cellcolor{pink}.56    & {Sleepy Joe - Biden}                        & .50           & \cellcolor{pink}{Boarding Schools - Uyghur}      & \cellcolor{pink}.55 \\             
             \midrule
                           & \cellcolor{pink}{Reeducation Camps - Camps}     & \cellcolor{pink}.62 & \cellcolor{green}{Isntreal - Israel}               & \cellcolor{green}.60    & \cellcolor{lightgray}{Anarkiddies - Tankies} & \cellcolor{lightgray}.63           & \cellcolor{lightgray}{Succ Dems - Socdems}       & \cellcolor{lightgray}.59 \\
              Conceptual   & \cellcolor{lightgray}{Succs - Socdems}          & \cellcolor{lightgray}.61 & \cellcolor{lime}{De-stalinization   - Khrushchev}  & \cellcolor{lime}.55    & \cellcolor{lime}{Trots - Tankies}            & \cellcolor{lime}.57           & \cellcolor{lightgray}{Anarchildren - Tankies}    & \cellcolor{lightgray}.53\\
                           & \cellcolor{lightgray}{Anarkids - Anarchists}    & \cellcolor{lightgray}.53 & \cellcolor{lime}{Destalinization   - Gorbachev}    & \cellcolor{lime}.55    & \cellcolor{cyan}{Zelensky - Trump}           & \cellcolor{cyan}.51           & \cellcolor{lime}{De-Stalinization - USSR}        & \cellcolor{lime}.53 \\
            Homomorphisms  & \cellcolor{lightgray}{Westoids - Libs}          & \cellcolor{lightgray}.53 & \cellcolor{lightgray}{Succ Dems - Socdems}         & \cellcolor{lightgray}.54    & \cellcolor{orange}{CGTN - CNN}               &  \cellcolor{orange}.51           & \cellcolor{lightgray}{Libshits - Libs}           & \cellcolor{lightgray}.52 \\
                           & \cellcolor{lightgray}{Libshits - Liberals}      & \cellcolor{lightgray}.53 & \cellcolor{lightgray}{Anarchildren - Anarchists}   & \cellcolor{lightgray}.53    & \cellcolor{lightgray}{Sucdems - Socdems}     & \cellcolor{lightgray}.50           & \cellcolor{lime}{Destalinization - Khrushchev}   & \cellcolor{lime}.52 \\
           
            \bottomrule
        \end{tabular}
        
      \end{adjustbox}
        \caption{Discovered misalignment and conceptually homomorphic word pairs that reflect polarized meanings from tankies and  r/communism, r/socialism, r/DemocraticSocialism, and r/Anarchism with their cosine similarities.
        We highlight pairs related to our main findings.
        Pink shows leaning towards accepting CCP narrative, lime shows leaning towards Stalinism, cyan shows leaning towards accepting Russian narrative in Ukraine, green shows anti-Zionist leaning, gray shows toxic tone, orange shows leaning towards state-sponsored media, and unhighlighted pairs are related to differences of tankies on US politics.}
        \label{tab:alignment_table}
    \end{table*}

    We then categorize pairs that reflect polarized meanings, informed by both the literature and our findings from other sections.
    We further validate our interpretations through qualitative and quantitative analyses on posts containing misaligned words, detailed in Appendix~\ref{sec:appendix_misalignment}.
    We exclude r/Marxism and r/IWW from our analysis due to their limited number of posts and vocabulary size (See Appendix~\ref{sec:appendix_data}, Figure~\ref{fig:total_number_of_posts}).
    We also exclude r/alltheleft because of the lack of literature regarding this community.
    Ultimately, we identify seven topics where tankies diverge from at least one of r/communism, r/socialism, r/DemocraticSocialism, or r/Anarchism: 
    1)~acceptance of CCP narratives, 
    2)~Stalinist leaning, 
    3)~acceptance of the Russian narrative in Ukraine, 
    4)~anti-Zionist leaning, 
    5)~toxic tone, 
    6)~leaning towards state-sponsored media, and
    7)~views on US politics.
    Below, we describe each of these in detail.

    \descr{Acceptance of CCP Narratives (Magenta).}
    Previously, we saw that the most popular topic tankies talk about is the Uyghur Genocide.
    Further analysis in Appendix~\ref{sec:ner_analysis} finds that China stands out as their most frequently mentioned named entity, surpassing other far-left communities. 
    While tankies may support the CCP's narrative due to China's AES status, as well as there being stories of r/GenZedong users attacking Uyghurs and promoting violence against them in the press~\cite{chow2022}, our previous analysis does not conclusively provide evidence on this.
    
    The first indication this is true from our misalignment analysis is tankies' use of the Chinese government's preferred nomenclature of the Communist Party of China (CPC)~\cite{li2012leadership,brown2017communist,peters2019chinese} over the more western term ``CCP.'' 
    Additional indication along these lines is how tankies are misaligned with respect to Uyghurs.
    There is substantial evidence that China detains Uyghur Muslims in so-called ``political re-education camps,'' known for extensive human rights abuses~\cite{hrwuyghur}, including evidence of genocide~\cite{stern2021genocide, turdush2021dossier, hoffpolicy}.
    This misalignment is demonstrated through ``boarding schools'' \& ``Uyghurs'' (i.e., that the Uyghur detention camps are merely typical schools), ``internment camps'' \& ``camps'', and ``reeducation camps'' \& ``camps'' pairs.
    
    Furthermore, tankies' discourse displays a shift in the characterization of individuals involved in civil resistance. 
    The CCP's historical framing of protesters as ``rioters''~\cite{fang1994riots} appears to align with this shift.
    As we check the top 10 most similar words to ``rioters'' in the embedding of tankies, we find Hong Kong and HK words.
    This is further supported by a ``Tiananmen Square'' \& ``riots'' misalignment pair.
    
    \descr{Stalinist Leaning (Lime).} Tankies lean towards Stalinism as outlined in Section~\ref{sec:background}.
    Yet, according to r/\hspace{0pt}InformedTankie, they mainly see themselves as Marxist-Leninists.
    So far, our previous analysis indicates a stronger focus on Stalin-related topics by tankies.
    They also mention Stalin more frequently than Lenin compared to other far-left communities, as detailed in Appendix~\ref{sec:ner_analysis}.
    
    Here, we identify several misaligning and conceptually homomorphic pairs suggesting a greater level of support for Stalinism among tankies.
    For example, we see misalignment in ``de-Stalinization'' \& ``Khrushchev'', ``destalinization'' \& ``Khrushchev'', ``Khrushchev'' \& ``Stalin'', and ``destalinization'' \& ``Gorbachev'' pairs.
    Nikita Khruschev, Stalin's successor as General Secretary of the CPSU (Communist Party of the Soviet Union), is known for his de-Stalinization efforts~\cite{filtzer1993khrushchev}.
    On the other hand, Mikhail Gorbachev is considered a successor of Khruschev's de-Stalinization efforts~\cite{jones2006dilemmas}.
    For r/Anarchism, ``de-Stalinization'' \& ``USSR'' is a conceptually homomorphic pair, which is meaningful given the history of Anarchists in Soviet Gulag Camps~\cite{confino1989varieties}.
    We also find ``trot'' \& ``Trotskyists'', ``trots'' \& ``Trotskyists'', ``trots'' \& ``tankies'', and ``Trotskyists'' \& ``Stalinists'' misalignment pairs.
    Interestingly, leading additional credence to tankies' Stalinist leaning, ``trot'' is a pejorative term used for Trotskyists, which was originally used by USSR-supporting elements during Stalin's reign~\cite{guardian2012}.
    
    \descr{Acceptance of the Russian Narrative in Ukraine (Cyan).}
    Tankies' support for the hardline Soviet era extends to Russia's current authoritarian regime's actions, especially when it comes to Russia's invasion of Ukraine~\cite{dutkiewicz2022why}.
    Although our previous findings in this section and Appendix~\ref{sec:ner_analysis} suggest that tankies show heightened interest in the Russian invasion of Ukraine compared to other far-left communities, we now provide evidence of their support.
    We find that tankies are more accepting of the chosen nomenclature from the self-proclaimed republics in Ukraine's Donbass region, Luhansk People's Republic (LPR) and Donetsk People's Republic (DPR), which were explicitly recognized by the Russian government and used as casus belli for the 2022 invasion of Ukraine~\cite{westfallparker2022}.
    We find that LPR and DPR are not in the vocabulary of r/DemocraticSocialism (and conceptually homomorphic pairs regarding LPR and DPR were off-topic). 
    This could be attributed to r/DemocraticSocialism's greater emphasis on US politics rather than on AES and the Russia-Ukraine conflict, as shown in Table~\ref{tab:others_top_NERS} in Appendix~\ref{sec:ner_analysis}. 
    Yet, a ``Zelensky'' \& ``Trump'' conceptually homomorphic pair between tankies and r/DemocraticSocialism suggests tankies' anti-Ukraine stance.

    \descr{Anti-Zionist Leaning (Green).}
    Since its establishment, the USSR actively pursued anti-Zionist efforts~\cite{segre2021anti}, and contemporary far-left anti-Zionism exhibits notable similarities to these campaigns~\cite{tabarovsky2019soviet}.  
    We see tankies lean more towards anti-Zionism from ``Zionist'' \& ``Israeli'', ``Zionists'' \& ``Jews'', ``Zionist'' \& ``Israel'', ``Zionism'' \& ``Israel'', and ``isntreal'' \& ``Israel'' pairs.
    These pairs show that tankies simplify the complexity and diversity of Jewish identity and also the establishment and development of the State of Israel.
    Appendix~\ref{sec:perspective} provides additional evidence to this effect, e.g., tankies have the largest proportion of posts with high identity attacks against Jews across all far-left communities. 
    
    \descr{Toxic Tone (Gray).}
    We find tankies tend to use pejorative terms to describe other ideologies in a toxic manner.
    E.g., anarkiddies, anarchildren, and anarkids instead of anarchists, succdems, succ dems,  and succs instead of socdems, and westoids and libshits when referring to liberals.
    We also find ``ultras'' \& ``tankies'' and ``radlibs'' \& ``tankies'' pairs.
    Considering that tankies is a pejorative term used by other ideologies, this finding means that tankies use ultras and radlibs in a pejorative way.
    As detailed in our toxicity analysis in Appendix~\ref{sec:perspective}, tankies have the highest percentage of identity attack and threat scores in their posts and rank second in toxicity, severe toxicity, profanity, and insult scores, only surpassed by r/alltheleft.
    
    \descr{Leaning Towards State-Sponsored Media (Orange).}
    We find ``RT'' \& ``NYT'' and ``CGTN'' \& ``CNN'' pairs between tankies and r/DemocraticSocialism.
    CGTN and RT are media organizations that receive direct support from their respective governments; CGTN is controlled by the Chinese government, while RT is backed by the Russian government~\cite{moore2022two,elswah2020anything}.
    As we find in our Domain Analysis in Appendix~\ref{sec:domain_main}, tankies also have the highest proportion of sharing state-sponsored or state-controlled news outlets from China and Russia.

    \descr{Views on US Politics (No Highlight).}
    We find that tankies frequently discuss Joe Biden, in the same way that r/\hspace{0pt}DemocraticSocialism do about Donald Trump.
    This tendency can be observed through the usage of phrases like ``Bidens'' \& ``Trumps'' and ``Sleepy Joe'' \& ``Biden.''
    Additionally, the anti-democratic leaning of tankies is evident from their discussions on topics, e.g., ``popularity contest'' \& ``elections.''

    \descr{Takeaways.}
    Analyzing the level of the usage of words between tankies and r/communism, r/socialism, r/\hspace{0pt}DemocraticSocialism, and r/Anarchism enabled us to quantify and expand the literature on the attributes of tankies (RQ3).
    We find that tankies show a stronger alignment with CCP narratives, Stalinism, Russia's stance on Ukraine, and exhibit a pronounced Anti-Zionist inclination.
    They often adopt a more toxic tone, using pejorative terms for other ideologies.
    Tankies also prefer state-backed media, especially compared to r/\hspace{0pt}DemocraticSocialism.
    Furthermore, tankies are less supportive of Joe Biden than r/DemocraticSocialism.
    By understanding these distinct tendencies, our findings contribute to the existing literature on tankies and the left in general, and offer a deeper understanding of their communication styles and ideological leanings within online communities.
    
    \section{Discussion \& Conclusion}
    
    In this paper, we presented a large-scale measurement of the left-wing extremist community known as ``tankies.''
    Although they have existed since the 1950s, when they supported hardline Soviet actions, more recently, the tankie ideology has shifted to support so-called Actually Existing Socialist (AES) countries, including China, USSR, and North Korea.
    To better understand tankies, we collected over 1.3M posts from 53K authors across six subreddits.

    Our analysis finds that the tankie community forms a distinct and cohesive group positioned at the periphery of the larger far-left spectrum on Reddit.
    We also observe that tankies gained increasing popularity within established far-left communities as their visibility and prominence grew.
    Further, notable differences in both the content and tone of tankie discussions compared to other far-left communities. 
    Our topic analysis reveals that tankies prioritize state-level political events and discussions surrounding authoritarian regimes over core left-wing issues like healthcare and housing. 
    Our lexical analysis shows consistent vocabulary misalignments and conceptual homomorphisms between tankies and other far-left communities, indicating tankies' divergent ideological focus. 
    Furthermore, we find tankies exhibit higher levels of toxicity, and they frequently share links from revisionist news sources.

    Our work constitutes the first large-scale, data-driven measurement of left-wing extremism on social media.
    The larger implication of our work is that the research community should (and must) broaden its focus when it comes to online extremism.
    While claims from the far-right (and more increasingly mainstream conservatives) characterizing leftist political movements like Antifa and Black Lives Matter as being dangerous extremists have mostly been unfounded, our results show that there \emph{is} a growing movement of far-left extremists operating online.
    And although the overwhelming majority of extremist actors online have been right-wing, the literature is quite clear on how state-sponsored actors have played both sides of the game, posing as leftists to foment dissent and polarization~\cite{bradshaw2022playing}.
    
    Consider the discussion around the October 7th, 2023 terrorist attack on Israel perpetrated by Hamas.
    At the time of this writing, there have been many prominent examples of left-leaning people diverging from the mainstream left.
    From politicians~\cite{Amiri_2023}, to news personalities~\cite{Knox_2023}, and even well standing members of the research community.
    These deviations from mainstream leftist perspectives among notable individuals emphasize why it is essential to measure and analyze far-left extremist movements, as we have done in this research.

    While it is unfair to characterize these dissenting individuals as extremists, it \emph{is} fair to note their alignment with tankies' views on, e.g., Zionism vs. the Israeli government (see Table~\ref{tab:alignment_table}).
    Further, it is important to note that many tankie or tankie aligned online influencers have been involved in spreading misinformation, hate speech, calls for violence, etc.
    
    By systematically measuring and analyzing tankies, we contribute to a more balanced and comprehensive understanding of online extremism across the political spectrum.
    Our methodologies can be applied to monitor and analyze other fringe or extremist groups across different platforms. 
    Our measurement pipeline can assist researchers and social media platforms in identifying and addressing the spread of harmful ideologies.
    Moreover, by comprehensively measuring this previously unexplored community, our findings on tankies can aid future research in identifying and comparing similar groups. 
    By doing so, researchers can build a more comprehensive picture of online extremism, contributing to the development of strategies to mitigate the spread of harmful ideologies.

    \section{Future Work}
    
    While we have mapped out the far-left and shown evidence that tankies are an extremist community within the far-left, our work raises at least as many questions as it answers.
    To that end, we have identified some particular lines of inquiry that we believe will be impactful moving forward.
    
    \descr{Comparison with the far-right.}
    The present work was focused almost exclusively on quantitatively mapping the far-left online with the goal of identifying extremists within it.
    Crucially, although we did not explore how the far-left compares with the far-right, many of our results indicate that left-wing extremists exhibit some of the same behaviors that make right-wing extremists such a danger to society.
    For example, the unwavering support for authoritarian governments (Russia in particular), coupled with our results that show how quickly the tankies community grew by siphoning off users from other far-left communities, implies a situation that looks a lot like the early days of the modern breed of far-right online extremists the research community has focused on so intently.

    Our results indicate that tankies share state-sponsored and revisionist misinformation just like right-wing extremists, but this is somewhat at odds with the substantial body of work that indicates right-wing users are more likely to engage with misinformation~\cite{davey2022far,zihiri2022qanon}.
    Explanations for this discrepancy range from an overall lack of knowledge and data on the far-left to implicit biases leading to methodological blind spots.
    Future work should thus revisit our understanding of misinformation and political polarization with an explicit attempt to include the impact of the far-left.

    \descr{Sentiment and topic dynamics.}
    Future work could more explicitly link sentiment measures to specific topics, highlighting how different communities not only discuss the same issues but also diverge in their attitudes toward them.  
    In parallel, examining tankies' engagement with non-political, everyday topics would help situate their extremist rhetoric within the broader texture of their online identity.
    
    \descr{What's in a name?}
    As noted previously, there has definitely been a drift from the original meaning of ``tankie'' to the current meaning.
    One (perhaps cynical) explanation for this might be that online spaces have enabled a more fundamental shift in extremism.
    For example, it remains unclear if tankies' online rhetoric translates to more traditionally explored aspects of leftist political movements like collective action, or if it is primarily a community engaged in extreme online contrarianism à la 4chan~\cite{hine2017kek}.
    
    At the same time, the presence of the ``tankie'' label itself suggests that we have a relatively limited understanding of the contemporary far-left's self-identification
    While ``tankie'' is a succinct, relatively easy to understand label that has a relatively clear etymology, and Figure~\ref{fig:atlas2_all_colored} shows that tankies form a clearly defined cluster of subreddits, the same figure also shows that the far-left is relatively diverse and there are many different ideologies with similar but meaningfully different labels (e.g., r/DemocraticSocialism and r/SocialDemocracy).
    
    Part of this is because the left has been historically (perhaps intrinsically) more fractured than the right.
    Another part of this is most certainly due to social media turbocharging niche communities' formation and evolution.
    We thus encourage future work that explores how other individual far-left online communities map to our understanding of those communities from the theoretical literature.
    We would be surprised if tankies are an outlier with respect to drifting from the original meaning of the label.

    \section{Limitations}
    
    Tankies are almost completely unexplored in measurement-driven research, and relatively unexplored in the broader literature when compared to more ``mainstream'' far-left ideologies like communism.
    This raises some concerns with respect to our interpretations and our dataset itself.
    While our results are definitely in-line with what little literature exists on the topic, the name ``tankies'' has absolutely drifted from its original meaning.
    We rely in large part on the tankie community on Reddit's self description of ``tankie,'' and although it aligns with recent use in the literature~\cite{petterson2020apostles}, it is worth recognizing that there is not a rich body of theoretical work for us to draw from when interpreting results.
    Our study focuses on six tankie subreddits, with a substantial portion of the data coming from r/GenZedong.
    This focus may inherently emphasize the perspectives and behaviors of a specific subset within the broader tankie community. 
    Additionally, smaller far-left subreddits are not included in our comparison with tankies.
    While these subreddits could offer alternative perspectives, we believe that the prominence and activity levels of the larger subreddits provide a strong and reliable foundation for studying the tankie community on Reddit.
    Additionally, our analysis is restricted to 2019-2022, which corresponds to the lifespan of the largest self-identified tankie subreddit on Reddit.
    This means we do not capture earlier phases of tankies on other platforms or the evolution of successor communities after 2022, which we leave for future work. 

    Despite the robustness of our methodology, it is important to acknowledge that our analysis is dependent on the accuracy of third-party tools and resources with publicly available licences are utilized.
    As such, potential limitations and sources of error should be considered, particularly in light of the findings of Hoffman et al.'s evaluation of such tools~\cite{hoffman2017evaluating}.
    Our results are ultimately bounded by the limitations of the tools we used, e.g., the Perspective API's potential inaccuracies and biases in toxicity detection~\cite{hosseini2017deceiving}, Word2Vec's lack of semantic understanding and limitations in capturing syntactic relationships~\cite{di2021considerations}, and BERTopic potentially generating more outliers than expected~\cite{egger2022topic}.
    That said, considering that our analysis is inline with the theoretical literature as well as empirical results from other online extremists, we have high confidence that our main findings hold.
    
    Finally, our dataset also has a notable lack of far-right communities.
    For example, there is no representation of fascism or alt-right communities.
    While the major explanation for this is relatively simple (Reddit has aggressively removed far-right extremists and some right-wing oriented subreddits like r/fascism are private), it is definitely a caveat when we compare tankies to right-wing extremism in general.

\bibliographystyle{ACM-Reference-Format}
%\bibliography{refs}

\newpage
\appendix

\section{Ethics}
To align our research with ethical standards, we follow established best practices, including refraining from further de-anonymizing any author~\cite{bailey2012menlo,rivers2014ethical}. 
Since our study uses only publicly accessible data and does not involve participant interactions, our institution does not classify it as human subjects research. 
Nevertheless, to prevent targeting any individual based on their comments used in our analysis, we paraphrase the text while maintaining its original tone and context when presenting examples.

\begin{figure*}[ht!]
	\centering
	\includegraphics[width=1\textwidth]{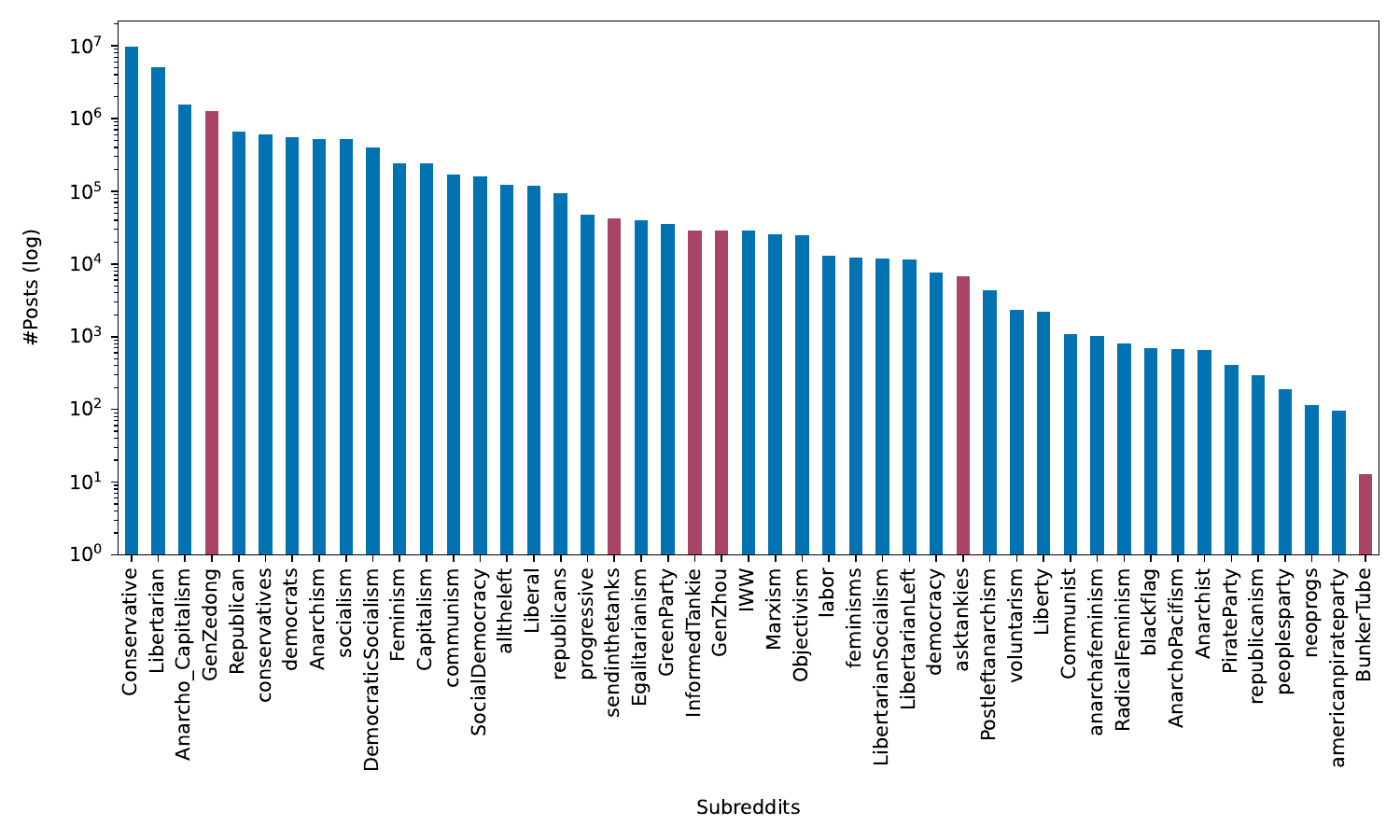}
	\caption{Total number of post from each community between July 15, 2019 and March 23, 2022. Tankie subreddits are specified with red.}
	\label{fig:total_number_of_posts}
\end{figure*}

\begin{table*}[t]
      \begin{center}
      
      \begin{tabular}{lrrrrrrrr}
          \toprule
          \multicolumn{1}{c}{Subreddit} & \multicolumn{1}{c}{\# Posts} & \multicolumn{1}{c}{\# Authors}  & \multicolumn{1}{c}{\begin{tabular}[c]{@{}c@{}}Min - Max\\ Date\end{tabular}}        & \multicolumn{1}{c}{In-Degree}                                             & \multicolumn{1}{c}{Out-Degree} & \multicolumn{1}{c}{PageRank} \\ 
          \midrule
          r/GenZedong                   & 1,273,277                   & 50,153                         & 07/2019 - 03/2022                                                                     &  7                                                                        & -                              & 0.00006 \\
          r/sendinthetanks              & 42,813                      & 5,603                          & 03/2020 - 03/2022                                                                     &  3                                                                        & -                              & 0.00003  \\
          r/InformedTankie              & 29,242                      & 3,697                          & 03/2020 - 03/2022                                                                     &  5                                                                        & 8                              & 0.00009     \\
          r/GenZhou                     & 28,606                      & 3,620                          & 12/2020 - 03/2022                                                                     &  3                                                                        & 10                             & 0.00004  \\
          r/asktankies                  & 6,678                       & 937                            & 01/2021 - 03/2022                                                                     &  2                                                                        & 2                              & 0.00003  \\
          r/BunkerTube                  & 12                          & 5                              & 11/2020 - 01/2022                                                                     &  1                                                                        & -                              & 0.00002  \\ 
      \bottomrule    
      \end{tabular}
      
      \end{center}
  \caption{General information and In-Degree, Out-Degree, and PageRank scores of the tankie subreddits from identified reference network. }
\label{tab:tankie_infos_appendix}
  \end{table*}

  \begin{table*}[th!]
    \centering
    \begin{adjustbox}{width=\textwidth}
    \begin{tabular}{lllllll}
        \toprule
        \multicolumn{1}{c}{}                    & \multicolumn{1}{c}{r/GenZedong}   & \multicolumn{1}{c}{r/sendinthetanks} & \multicolumn{1}{c}{r/InformedTankie} & \multicolumn{1}{c}{r/GenZhou} & \multicolumn{1}{c}{r/asktankies} & \multicolumn{1}{c}{r/Bunkertube} \\ 
        \midrule
        \multirow{7}{*}{Subreddits referred by}   &  CommunismMemes                   & InformedTankie                      &   CPUSA                               &  asktankies                  & InformedTankie                    & InformedTankie                                                 \\ 
                                                &  Sino                             & ComradesInAmerica                   &   ComradesInAmerica                   &  ItaliaRossa                 & ItaliaRossa                       &                                                  \\ 
                                                &  GenHoChiMinh                     & Enough\_VDS\_Spam                   &   BasedLibrary                        &  GenHoChiMinh                &                                   &                                              \\
                                                &  ChunghwaMinkuo                   &                                     &   Sino                     &                              &                                   &                  \\                    
                                                &  WorkersStrikeBack                &                                     &                                   &                              &                                   &                                   \\ 
                                                &  LeftieZ                          &                                     &                                       &                              &                                   &                                   \\
                                                &  InformedTankie                   &                                     &                                       &                              &                                   &                                   \\
          \midrule
        \multirow{10}{*}{Subreddits referred to}   &                                   &                                    & GenZedong                             & PoliticalDiscussion          & InformedTankie                   &            \\
                                                &                                   &                                     & sendinthetanks                        & nato                         & GenZhou                           &            \\
                                                &                                   &                                     & 52BooksForCommunists                  & communism101                 &                                   &            \\
                                                &                                   &                                     & CPUSA                                 & answers                      &                                   &            \\
                                                &                                   &                                     & ComradesInAmerica                     & DebateCommunism              &                                   &             \\
                                                &                                   &                                     & BunkerTube                            & AskARussian                  &                                   &             \\
                                                &                                   &                                     & asktankies                            & tankiejerk                   &                                   &             \\
                                                &                                   &                                     & BasedLibrary                          & askaconservative             &                                   &             \\
                                                &                                   &                                     &                                       & socialism\_101               &                                   &             \\
                                                &                                   &                                     &                                       & centrist                     &                                   &             \\
    
        \bottomrule    
    \end{tabular}
  \end{adjustbox}
    \caption{Subreddits that are linked to tankie subreddits in the identified reference network.}
  \label{tab:tankies_network_subs}
    \end{table*}
  
    \begin{table*}[th!]
      \centering
      \begin{adjustbox}{width=\textwidth}
      \begin{tabular}{ll}
          \toprule
          \multicolumn{1}{c}{Subreddit} & \multicolumn{1}{c}{Definition} \\ 
          \midrule
          r/52BooksForCommunists          & A book discussion page specifically for communists.                  \\
          r/answers                       & A Q\&A subreddit for the Reddit users.                  \\
          r/askaconservative              & A Q\&A subreddit that Reddit users can ask (mostly political) questions to be aswered by conservatives.                   \\       
          r/AskARussian                   & A Q\&A subreddit that Reddit users can ask questions to be aswered by Russians.                      \\
          r/BasedLibrary                  & A discussion page specifically for books, videos, and images about communism.                       \\
          r/centrist                      & A subreddit for the Reddit users who sees themselves at the center of the political spectrum.                   \\
          r/ChunghwaMinkuo                & A subreddit for anyone that supports Taiwan.                    \\    
          r/communism101                  & A subreddit dedicated for teaching communism to the Reddit users.                       \\
          r/CommunismMemes                & A place for sharing memes related to communism.                         \\
          r/ComradesInAmerica             & A subreddit for Marxist, Leninists, and Maoists in the US.                           \\                              
          r/CPUSA                         & An unoficcial subreddit dedicated to the Communist Party of USA.                \\
          r/DebateCommunism               & A debate subreddit oriented around discussions related to communism and socialism.                  \\        
          r/Enough\_VDS\_Spam             & A subreddit dedicated to defend Vaush, a libertarian socialist video streamer.               \\
          r/GenHoChiMinh                  & A Vietnamese Marxist/Leninist subreddit dedicated for the former president of Vietnam, Ho Chi Minh.           \\
          r/ItaliaRossa                   & A subreddit for the Italian Marxist/Leninists.         \\
          r/LeftieZ                       & A communist leaning leftist subreddit centered around young Reddit users.     \\
          r/nato                          & A subreddit for discussing and sharing news related to NATO (North Atlantic Treaty Organization).  \\
          r/PoliticalDiscussion           & A subreddit that Reddit users can discuss about politics.    \\             
          r/Sino                          & A pro-China subreddit for any content relateed to China and Chinese.  \\
          r/socialism\_101                & A subreddit dedicated for teaching socialism to the Reddit users.    \\        
          r/tankiejerk                    & A subreddit for criticizing Tankies from a leftist perspective.      \\  
          r/WorkersStrikeBack             & A socialist subreddit dedicated to support workers and their rights.        \\                         
          \bottomrule    
      \end{tabular}
    \end{adjustbox}
      \caption{Definition of the subreddits that are linked to tankie subreddits in the identified reference network. }
    \label{tab:network_subs_definition}
      \end{table*}

      \begin{figure*}[t]
        \centering
        \includegraphics[width=1\textwidth]{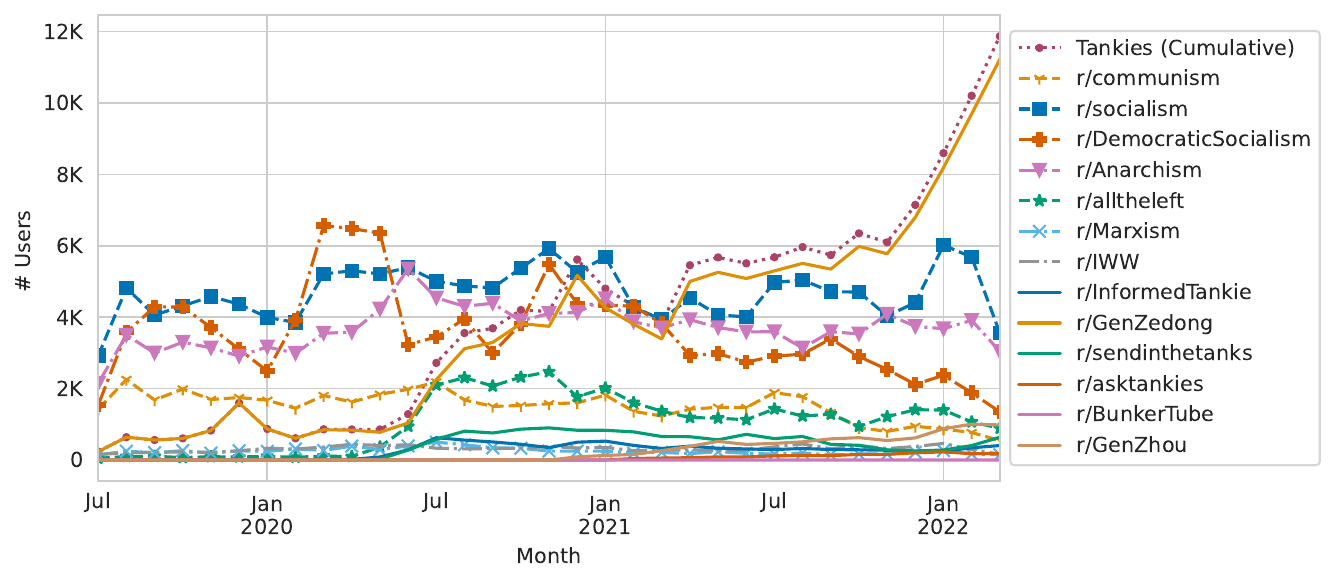}
        \caption{Monthly active user counts of communities in the far-left community. Tankies outgrow the monthly active user counts of any other far-left community starting from April 2021.
                    }
        \label{fig:appendix_monthly_user_counts}
      \end{figure*}

\section{MEASURING TANKIE ACTIVITY}
\label{sec:appendix_data}

\subsection{Tankie Subreddits}

\descr{r/GenZedong} is the most popular tankie subreddit we find, with nearly 1.3M posts created by 50K authors.
With a first post date in July 2019, it is also the oldest tankie subreddit we discover.
r/GenZedong itself does not link to any subreddit, and thus has zero out-degree in our reference network graph.
On the other hand, it has a 0.00006 PageRank score, making it the second-highest ranked tankie subreddit.
As additional evidence that tankies' are indeed supporters of AES countries like China, we note that a Taiwanese nationalist subreddit links to r/GenZedong, describing its members as ``Communist Bandits.''
Due to spreading misinformation, r/GenZedong was quarantined by Reddit on March 23, 2022~\cite{redditgenzedong}.
A Previous study~\cite{norambuena2022characterizing} regarding the 2021 Cuban Protests found that r/GenZedong supports the Cuban regime while condemning those who oppose it.

\descr{r/sendinthetanks}, with 42,813 posts from 5,603 authors, and an earliest post date of March 2020, is the second most popular tankie subreddit in our dataset.
Besides leftist subreddits, we find r/Enough\_VDS\_Spam (a subreddit dedicated to defending Vaush, a libertarian socialist streamer~\cite{reutersVaush}) links to this subreddit under the heading ``Sources of Cringe.''
This is an early indication of polarization between tankies and some far-left communities.
Despite its popularity, r/sendinthetanks has the second lowest PageRank score of our tankie subreddits.
One possible reason for this could be the content and focus of the subreddit.
While other tankie subreddits may be more centered on promoting certain ideologies or agendas, r/sendinthetanks may be more geared towards discussion and debate among its members. 
This could lead to a lower PageRank score, as it is not as actively linked to or referenced by other subreddits.

\descr{r/InformedTankie}, the third most popular tankie subreddit, has 29,242 posts from 3,697 authors dating back to March 2020.
r/InformedTankie's banner defines ``tankie'' in line with Petterson's description (See Background):

\begin{quote}
    For those who aren't aware of what ``tankie'' means, its context has culturally changed in the online left from meaning people who support the USSR sending tanks into Hungary to crush the fascist coup, to being an umbrella term for communists in general, especially Marxist-Leninists who defend and/or support the Actually Existing Socialist (AES) countries.
  \end{quote}

It is listed by r/Sino as a subreddit that can be crossposted from, likely indicating a pro-China bias among its members.
Our findings show that r/InformedTankie has the highest PageRank score, as well as the second highest in-degree and out-degree scores of tankie subreddits, which supports our previous hypothesis that it serves as a hub for the tankie community.
We also find that r/InformedTankie has the highest number of outbound links directed toward other tankie subreddits (4).

\descr{r/GenZhou} is the fourth most popular tankie subreddit, with 28,606 posts from 3,620 authors since December 2020.
r/GenZhou refers not only to leftist subreddits, but also subreddits that are ideologically \emph{opposed} to tankies (r/tankiejerk, r/askaconservative, r/centrist, and r/nato) as well as debate subreddits (r/answers and r/PoliticalDiscussion).
One explanation might be that r/GenZhou could be pointing these communities to their user base to call for support in certain discussions.
We also find r/GenZhou referring to r/AskARussian, which might reflect a pro-Russian leaning.
Reddit banned r/GenZhou on April 4, 2022~\cite{redditgenzhou} for ``creating or repurposing a sub to reconstitute or serve the same objective as a previously banned or quarantined subreddit'' (i.e., ban evasion).

\descr{r/asktankies} is the fifth most popular tankie subreddit with 6,678 posts from 937 authors since January 2021.
It is a Q\&A subreddit in which questions are answered by tankie users.

\descr{r/Bunkertube} is the least popular tankie subreddit, with only 12 posts from 5 authors since November 2020.
It has the lowest in-degree, out-degree, and PageRank scores among tankie subreddits.

  \subsection{Network Metrics}

\begin{itemize}
  \item \textbf{In-Degree}: Number of edges pointing to a node from other nodes.
  \item \textbf{Out-Degree}: Number of edges a node pointing to other nodes.
  \item \textbf{PageRank}: Quantifies the structural importance of a node by ranking it based on the relative importance of its inbound edges.
\end{itemize}

\begin{table*}[t]
  \centering
  \begin{tabular}{rlrr|rlrr}
      \toprule
      \multicolumn{1}{c}{Rank} & \multicolumn{1}{c}{Named Entity} & \multicolumn{1}{c}{\#Posts} & \multicolumn{1}{c}{\%} & \multicolumn{1}{c}{Rank} & \multicolumn{1}{c}{Named Entity} & \multicolumn{1}{c}{\#Posts} & \multicolumn{1}{c}{\%} \\ 
      \midrule
      1                        & China            &   81,115    & 5.87  & 11                       & Ukraine          &   13,606    & 0.98      \\
      2                        & US               &   48,809    & 3.53  &12                       & Russian          &   12,622    & 0.91       \\
      3                        & Chinese          &   30,985    & 2.24  &13                       & Two              &   11,490    & 0.83           \\
      4                        & One              &   25,123    & 1.82  &14                       & CIA              &   10,801    & 0.78      \\
      5                        & Russia           &   23,183    & 1.68  &15                       & Nazis            &   10,679    & 0.77      \\
      6                        & American         &   21,860    & 1.58  &16                       & Americans        &   10,218    & 0.74       \\
      7                        & First            &   20,988    & 1.52  &17                       & USA              &   9,650     & 0.69     \\
      8                        & America          &   18,943    & 1.37  &18                       & Lenin            &   8,996     & 0.65        \\
      9                        & Stalin           &   14,529    & 1.05  &19                       & Mao              &   8,780     & 0.63        \\
      10                       & USSR             &   13,611    & 0.98  &20                       & Today            &   7,894     & 0.57      \\   
      
      \bottomrule    
  \end{tabular}
  \caption{Top 20 named entities of tankies.}
\label{tab:tankies_top_20_NER}
  \end{table*}

  \begin{table*}[h]
    \centering
    \begin{adjustbox}{width=\textwidth}
      \begin{tabular}{lrrlrrlrrlrr}
          \toprule
          \multicolumn{3}{c|}{r/communism}         & \multicolumn{3}{c|}{r/socialism}               & \multicolumn{3}{c|}{r/DemocraticSocialism}         & \multicolumn{3}{c}{r/Anarchism}            \\
          \midrule
          Named Entity  & \#Posts & \%                 & Named Entity  & \#Posts & \%                  & Named Entity  & \#Posts & \%                    & Named Entity  & \#Posts & \%                 \\                                                 
          \midrule
          China         & 7,157  & 4.15               & US         & 17,172 & 3.33                  & Biden        & 13,821  & 3.45                 & One         & 13,263 & 2.56               \\
          US            & 5,635  & 3.27               & One        & 12,362 & 2.40                  & Trump        & 11,000  & 2.74                 & First       & 11,879 & 2.29               \\
          One           & 4,162  & 2.41               & First      & 11,356 & 2.20                  & Bernie       & 10,281  & 2.56                 & US          & 10,038 & 1.93               \\
          First         & 3,959  & 2.29               & China      & 10,491 & 2.03                  & US           & 8,616   & 2.15                 & Two         & 6,737  & 1.30               \\
          USSR          & 2,942  & 1.70               & American   & 7,573  & 1.47                  & One          & 8,564   & 2.13                 & American    & 5,012  & 0.96               \\
          Marxist       & 2,895  & 1.68               & Two          & 6,740  & 1.30                & First         & 7,200   & 1.79                & China         & 4,770   & 0.92 \\
          Chinese       & 2,649  & 1.53               & America      & 6,578  & 1.27                & democrats     & 7,116   & 1.77                & Today         & 4,298   & 0.83 \\
          American      & 2,597  & 1.50               & USSR         & 5,478  & 1.06                & Two           & 5,916   & 1.47                & America       & 4,164   & 0.80 \\
          Marx          & 2,333  & 1.35               & Marx         & 5,470  & 1.06                & America       & 5,881   & 1.46                & 2             & 3,325   & 0.64 \\
          Two           & 2,331  & 1.35               & Today        & 5,096  & 0.98                & Republicans   & 5,758   & 1.43                & Second        & 2,937   & 0.56 \\
          Two           & 2,331  & 1.35               & Today      & 5,096  & 0.98                      & Republicans  & 5,758   & 1.43                     & Second      & 2,937  & 0.56                            \\
          Stalin        & 2,295  & 1.33               & Russia     & 4,675  & 0.90                      & American     & 5,316   & 1.32                     & 1           & 2,864  & 0.55                               \\              
          Lenin         & 2,238  & 1.29               & Cuba       & 4,457  & 0.86                      & Americans    & 3,720   & 0.92                     & Reddit      & 2,858  & 0.55                                 \\
          Today         & 1,956  & 1.13               & Marxist    & 4,283  & 0.83                      & Republican   & 3,365   & 0.84                     & Trump       & 2,846  & 0.54                          \\
          Cuba          & 1,837  & 1.06               & Socialists & 4,072  & 0.79                      & DNC          & 3,009   & 0.75                     & Russia      & 2,758  & 0.53                              \\
          America       & 1,761  & 1.02               & Lenin      & 4,065  & 0.78                      & 2            & 2,928   & 0.73                     & Marx        & 2,519  & 0.48                                       \\  
          Russia        & 1,653  & 0.95               & Stalin     & 3,701  & 0.71                      & Democratic   & 2,922   & 0.72                     & Nazis       & 2,469  & 0.47                                 \\ 
          Mao           & 1,592  & 0.92               & Trump      & 3,666  & 0.71                      & Democrat     & 2,760   & 0.68                     & Anarchists  & 2,298  & 0.44                              \\ 
          Soviet        & 1,532  & 0.88               & 2          & 3,600  & 0.69                      & Today        & 2,481   & 0.61                     & 3           & 2,224  & 0.42                                  \\ 
          Soviet Union  & 1,230  & 0.71               & Chinese    & 3,322  & 0.64                      & 1            & 2,331   & 0.58                     & USSR        & 2,007  & 0.38                                \\
          Communist     & 1,224  & 0.71               & Second     & 3,258  & 0.63                      & Obama        & 2,316   & 0.57                     & Biden       & 1,947  & 0.37                                  \\
          \midrule
          \multicolumn{3}{c|}{r/alltheleft}         & \multicolumn{3}{c|}{r/Marxism}               & \multicolumn{3}{c|}{r/IWW}            \\
          \midrule
          Named Entity  & \#Posts & \%                 & Named Entity  & \#Posts & \%                  & Named Entity  & \#Posts & \%                 \\                                                 
          \midrule
          US              & 2,328 & 1.91              & Marx       & 3,297 & 12.91               & IWW         & 2,932    & 10.2               \\
          Biden           & 2,188 & 1.79              & Marxist    & 1,817 & 7.11                & One         & 903      & 3.16               \\
          One             & 2,016 & 1.65              & One        & 1,203 & 4.71                & First       & 678      & 2.37               \\
          Trump           & 1,947 & 1.60              & First      & 1,169 & 4.57                & US          & 439      & 1.53               \\
          First           & 1,744 & 1.43              & Two        & 754   & 2.95                & Two         & 387      & 1.35               \\
          Democrats       & 1,430 & 1.17                    & Marxists & 730   & 2.85         & Solidarity &  354      & 1.23               \\          
          American        & 1,394 & 1.14                    & Lenin    & 670   & 2.62         & Today      &  273      & 0.95               \\
          America         & 1,272 & 1.04                    & US       & 588   & 2.30         & 1          &  219      & 0.76               \\
          Two             & 1,187 & 0.97                    & China    & 566   & 2.21         & 2          &  217      & 0.75              \\
          100\%           & 1,168 & 0.95                    & Today    & 547   & 2.14         & American   &  193      & 0.67              \\
          China           & 994   & 0.81                    & USSR     & 475   & 1.86         & GEB        &  161      & 0.56                 \\
          Republicans     & 942   & 0.77                    & Engels   & 474   & 1.85         & Second     &  159      & 0.55              \\
          Second          & 780   & 0.64                    & Stalin   & 452   & 1.77         & 3          &  155      & 0.54                \\
          Bernie          & 748   & 0.61                    & 1        & 447   & 1.75         & Wobblies   &  151      & 0.52                \\
          Americans       & 680   & 0.55                    & 2        & 389   & 1.52         & Amazon     &  145      & 0.50                   \\ 
          Obama           & 623   & 0.51                    & Second   & 337   & 1.32         & DSA        &  131      & 0.45                 \\
          Today           & 621   & 0.51                    & 3        & 284   & 1.11         & Reddit     &  130      & 0.45               \\
          shitpisscum1312 & 595   & 0.48                    & American & 282   & 1.10         & America    &  128      & 0.44                  \\     
          Fucking cooldownbot & 595 & 0.48                  & Russia   & 223   & 0.87         & NLRB       &  124      & 0.43               \\
          Russia          & 572   & 0.47                    & German   & 219   & 0.85         & Biden      &  121      & 0.42                 \\                       
          \bottomrule
      \end{tabular}
   \end{adjustbox}
    \caption{Top 20 named entities of far-left communities other than tankies}
    \label{tab:others_top_NERS}
  \end{table*}

\subsection{User Migrations Over Time}
\label{sec:migration}

During the second month after r/GenZedong was founded, 32.4\% of the tankies user base came from other far-left communities.
On average, 27.7\% of tankies MAU came from other far-left communities during the first period.
However, during the second period, there was a clear decrease in migrations from other far-left communities to tankies, with a mean MRD of 14.9\%.
The highest migrations to tankies come from r/communism and r/socialism for most of the months in our dataset. 
During the first period, the mean MRDs from these communities to tankies are 13\%.
The second period also saw a decrease in MRDs from these communities, with mean MRDs of 4.8\% and 7.4\% respectively.

We find strong negative correlations between the monthly user counts of tankies and MRDs to tankies from r/\hspace{0pt}DemocraticSocialism ($\rho = -0.78$), r/Anarchism ($\rho = -0.78$), r/communism ($\rho = -0.74$), r/socialism ($\rho = -0.72$), and r/Marxism ($\rho = -0.72$).
We also find a moderate negative correlation for r/IWW ($\rho = -0.67$).
In contrast, we find a moderate positive correlation for r/alltheleft ($\rho = 0.48$).
We do not observe any trend for this community's MRDs to tankies, and we find decreasing trends for MRDs from other far-left communities to tankies when applying the Mann-Kendall test.

For the MRSs of other far-left communities to tankies, a different pattern emerges.
In the second time period, we find that the MRSs to tankies from these communities are higher.
Instead of r/socialism, we find r/Marxism and r/communism have the highest MRSs to tankies for most of the months in our dataset's timeline.
The mean MRSs for r/communism and r/Marxism are 6.4\% and 7.5\% in the first period, but they increase to 16\% and 13.3\% in the second period.
This means that over 1 in 10 users from r/communism and r/Marxism continuously migrated to tankies during this period.
This trend is also seen for r/alltheleft, except for three months.
Our findings show that r/DemocraticSocialism and r/Anarchism have the lowest migrations to tankies, with mean migrations of 1\% and 1.5\% in the first period, and 2.4\% and 2.6\% in the second period, respectively.

In contrast to MRDs, we find positive strong correlations between the monthly user counts of tankies and MRSs from other far-left communities.
r/socialism ($\rho = 0.93$), r/alltheleft ($\rho = 0.91$), r/DemocraticSocialism ($\rho = 0.89$), and r/IWW ($\rho = 0.72$) have strong correlations, while r/communism ($\rho = 0.68$), r/Anarchism ($\rho = 0.62$), and r/Marxism ($\rho = 0.50$) have moderate positive correlations.
We also find increasing trends for MRSs from each far-left community to tankies, except for r/Anarchism.

We find all correlations are significant, with $p < 0.01$ after adjustment for multiple testing using the Benjamini-Hochberg method~\cite{benjamini1995controlling}.

\section{Who are tankies talking about?}
\label{sec:ner_analysis}
Although topic analysis provides us meaningful insights about the differences in the focus of discussion between tankies and other far-left communities, we are limited to the information we can collect from the words in the topic, and some topics contain multiple, opposing actors; e.g., we can not understand if tankies talk more about Russia or Ukraine in the related topic.
Therefore, we examine named entities used by far-left communities to gain a deeper understanding of the differences in their focus.

We extract named entities using the \textit{en\_core\_web\_lg} model of the SpaCy library~\cite{spacy}, which  has previously been used in studies on social media platforms and news articles~\cite{papasavva2020raiders,aliapoulios2021gospel}.

\descr{Results.}
In Table~\ref{tab:tankies_top_20_NER}, we present the top 20 named entities, their frequency, and the percent of posts they appear in for tankies.
Table~\ref{tab:others_top_NERS} shows the same information for other far-left subreddits. 
The most popular named entities tankies mention are related to China (China, Chinese, Mao), the United States (US, American, America, Americans, CIA, USA), the USSR (Stalin, USSR, Lenin), Russia, Ukraine, Nazis, and some numeric and date related named entities.
We also observe similar entities being discussed on r/communism, r/socialism, r/Marxism, and r/Anarchism, with only small differences. 
Notably, Ukraine is only present in the most popular named entities of tankies.
Consistent with our findings in Section~\ref{sec:what_do_tankies_talk_about}, this likely indicates a focus on Russia-Ukraine relations and specifically the 2022 Russian invasion of Ukraine.

When looking at the specific named entities, we see that tankies mention China at a higher rate than other far-left subreddits.
Additionally, tankies mention Stalin nearly twice as often as Lenin.
This is interesting because we find other far-left communities mention Lenin in similar or higher proportions compared to Stalin.
As outlined in Section~\ref{sec:what_do_tankies_talk_about}, tankies tend to focus more on topics related to China and Stalin compared to other far-left communities. 
These disparities show that the inclination of tankies towards China and Stalinism is a defining characteristic, which aligns well with Petterson's description.

Finally, we compare the similarity of the most popular named entities of tankies and other far-left communities by using Rank Biased Overlap~\cite{webber2010similarity}, which measures the similarity of two indefinite ranked lists by giving weight to top N ranks, where the ranked lists do not need to be conjoint.
The results indicate that r/communism is the most similar to tankies in terms of named entity rankings, with a score of 0.74, followed by r/socialism (0.70), r/Anarchism (0.62), r/alltheleft (0.55), r/Marxism (0.53), r/DemocraticSocialism (0.39), and r/IWW (0.38), respectively.

\descr{Takeaways.}
The results of our analysis reveal that the most popular named entities mentioned by tankies are related to China, the United States, the USSR, Russia, Ukraine, and Nazis.
Additionally, we find that tankies mention China at a higher rate and Stalin nearly twice as often as Lenin, compared to other far-left communities. 
Our findings demonstrate that the preference for China and Stalinism is a defining characteristic of tankies.
Furthermore, our analysis reveals that Ukraine is among the most popular named entities of the tankies community, and not among the most popular of the other communities.
This suggests that the topic of Russia-Ukraine relations is more relevant or important to tankies, or that this community's members engage in more discussion on this topic than those of other far-left communities.

\section{What do tankies talk about?}
\label{sec:appendix_topic}

\subsection{ Model Training}
We leverage BERTopic~\cite{grootendorst2022bertopic}, a technique that uses pre-trained, transformer-based language models to create document embeddings, applies dimensionality reduction with UMAP~\cite{mcinnes2018umap}, hierarchically clusters these embeddings with HDBSCAN~\cite{mcinnes2017hdbscan}, and forms topic representations with class-based TF-IDF.
Unlike traditional methods (e.g., LDA), BERTopic focuses on \emph{documents} rather than words, considering word context and semantics.
Previous research~\cite{egger2022topic} found that it performs better in terms of topic separation than LDA and another embedding-based topic modeling approach, Top2Vec~\cite{angelov2020top2vec}.
After removing hyperlinks, and posts from ``Automoderator'' (a bot which handles a variety of automated moderation tasks), we train our model with default parameters.

\section{What are their lexical differences?}
\label{sec:appendix_misalignment}
If a word from a source community corresponds to a different word in a target community, we consider the words to be ``misaligned.'' 
We can further discover \emph{conceptual homomorphisms} between communities, enabling us to identify equivalent words in the target community for unique words or slang used in the source community.
This is accomplished by projecting the words from the source community's vocabulary that are not in the vocabulary of the target community to target community's vocabulary.
This allows us to get a much deeper understanding of where tankies sit with respect to the rest of the far-left.

\begin{figure}[t]
  \centering
  \begin{subfigure}[b]{0.45\textwidth}
      \centering
      \includegraphics[width=\linewidth]{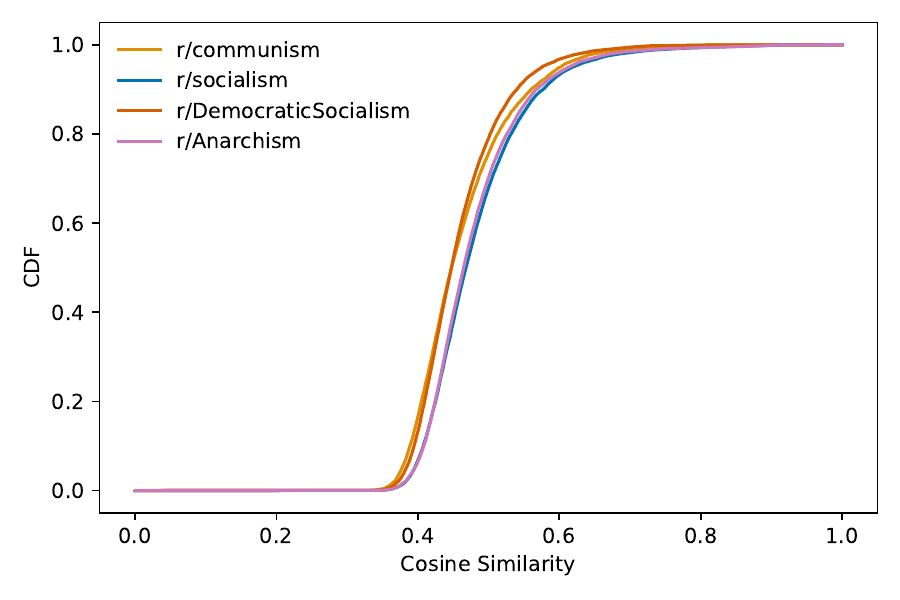}
      \caption{Misalignments}
      \label{fig:misalignment_cosine_cdf}
  \end{subfigure}
  \begin{subfigure}[b]{0.45\textwidth}
      \centering
      \includegraphics[width=\linewidth]{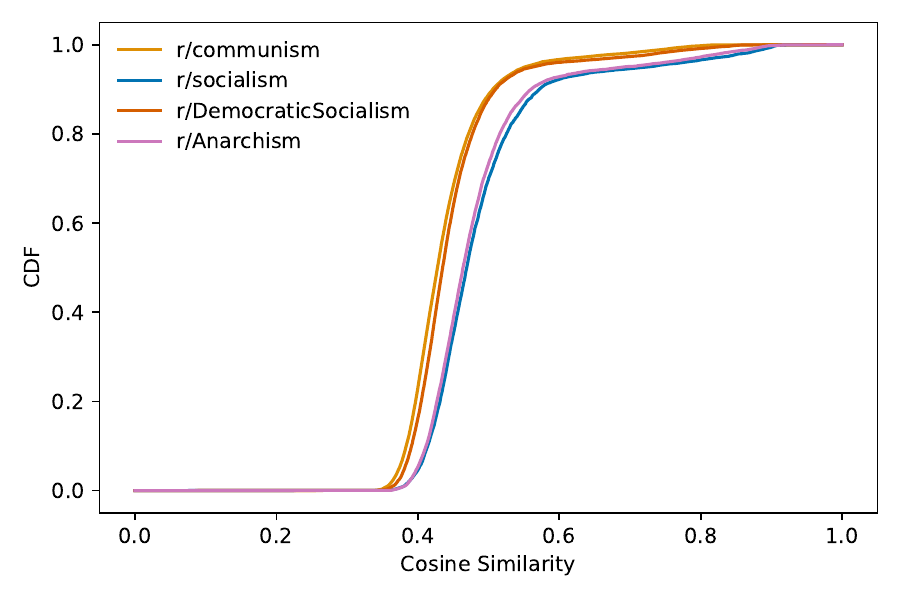}
      \caption{onceptual homomorphisms}
      \label{fig:conceptual_cosine_cdf}
  \end{subfigure}
  
     \caption{CDF of cosine similarities of misaligned and conceptually homomorphic pairs between tankies and r/communism, r/socialism, r/DemocraticSocialism, and r/Anarchism.}
     \label{fig:cosine_similarity_cdfs}
\end{figure}

\begin{table*}[htp!]
  \begin{scriptsize}
      \begin{center}
      
      \begin{tabular}{lrrrrr}
          \toprule
          \multicolumn{1}{c}{Community} & \multicolumn{1}{c}{\#Posts} & \multicolumn{1}{c}{Mean} & \multicolumn{1}{c}{Median}  & \multicolumn{1}{c}{\begin{tabular}[c]{@{}c@{}}Standard\\ Deviation\end{tabular}}        & \multicolumn{1}{c}{\%High}                                          \\ 
          \midrule
          \multicolumn{6}{c}{CCP - CPC}  \\
          \midrule                                                                                                                                                                            
          Tankies~\tiny\textbf{(cpc)}                 & 7,322                   & 0.14                         & 0.05                 &  0.19            & 0.86                                                            \\
          r/communism~\tiny\textbf{(ccp)}             & 533                     & 0.13                         & 0.06                 &  0.17            & 0.94                                               \\
          r/socialism~\tiny\textbf{(ccp)}             & 807                     & 0.15                         & 0.07                 &  0.19            & 1.61                                                 \\
          r/DemocraticSocialism~\tiny\textbf{(ccp)}   & 227                     & 0.11                         & 0.02                 &  0.19            & 0.88                                                   \\
          r/Anarchism~\tiny\textbf{(ccp)}             & 774                     & 0.23                         & 0.13                 &  0.23            & 4.01                                                    \\
          \midrule
          \multicolumn{6}{c}{Uyghur Genocide  \#1}  \\
          \midrule                                                                                                                                                                            
          Tankies~\tiny\textbf{(boarding schools, reeducation camps,internment camps)}       & 380                    & 0.23                         & 0.17                 &  0.19            & 2.13                                                            \\
          r/communism~\tiny\textbf{(uyghurs)}             & 100                    & 0.19                         & 0.13                 &  0.18            & 0.00                                               \\
          r/socialism~\tiny\textbf{(uighurs)}             & 171                  & 0.20                         & 0.14                 &  0.18            & 0.58                                                 \\
          r/DemocraticSocialism~\tiny\textbf{(uyghur)}    & 44                     & 0.08                         & 0.02                 &  0.12            & 0.00                                                   \\
          r/Anarchism~\tiny\textbf{(uyghurs)}             & 162                    & 0.22                         & 0.14                 &  0.20            & 0.61                                                    \\
          \midrule
          \multicolumn{6}{c}{Uyghur Genocide  \#2}  \\
          \midrule                                                                                                                                                                            
          Tankies~\tiny\textbf{(boarding schools, reeducation camps,internment camps)}       & 380                    & 0.23                         & 0.17                 &  0.19            & 2.13                                                            \\
          r/communism~\tiny\textbf{(camps)}             & 497                    & 0.20                         & 0.13                 &  0.18            & 0.60                                               \\
          r/socialism~\tiny\textbf{(camps)}             & 1,382                  & 0.22                         & 0.16                 &  0.18            & 0.72                                                 \\
          \midrule
          \multicolumn{6}{c}{Rioters - Protesters}  \\
          \midrule                                                                                                                                                                            
          Tankies~\tiny\textbf{(rioters)}       & 490                    & 0.26                         & 0.19                 &  0.22            & 2.65                                                            \\
          r/communism~\tiny\textbf{(protesters)}             & 314                    & 0.17                         & 0.09                 &  0.18            & 0.96                                               \\
          r/socialism~\tiny\textbf{(protesters)}             & 1,139                  & 0.17                         & 0.10                 &  0.18            & 0.96                                                \\
          r/DemocraticSocialism~\tiny\textbf{(protesters)}    & 428                     & 0.10                         & 0.02                 &  0.15            & 0.70                                                   \\
          r/Anarchism~\tiny\textbf{(protestors)}             & 1,025                    & 0.23                         & 0.14                 &  0.22            & 1.66                                                    \\
          \midrule
          \multicolumn{6}{c}{Tiananmen Square - Riots}  \\
          \midrule                                                                                                                                                                            
          Tankies~\tiny\textbf{(tiananmen square)}       & 880                    & 0.19                         & 0.09                 &  0.20            &0.87                                                            \\
          r/communism~\tiny\textbf{(riots)}             & 172                    & 0.18                         & 0.09                 &  0.18            & 0.0                                               \\
          r/socialism~\tiny\textbf{(riots)}             & 561                  & 0.17                         & 0.09                 &  0.17            & 0.53                                                \\
          r/DemocraticSocialism~\tiny\textbf{(riots)}    & 363                     & 0.09                         & 0.01                 &  0.15            & 0.53                                                   \\
          r/Anarchism~\tiny\textbf{(riots)}             & 923                    & 0.19                         & 0.09                 &  0.20            & 0.97                                                    \\
          \midrule
          \multicolumn{6}{c}{Stalinist Leaning \#1}  \\
          \midrule                                                                                                                                                                            
          Tankies~\tiny\textbf{(trots, de(-)stalinization, khrushchev, trot, trotskyists)}   & 4,044                   & 0.18                         & 0.08                 &  0.20            & 1.61                                                            \\
          r/communism~\tiny\textbf{(trotskyist, khrushchev)}                                 & 522                     & 0.11                         & 0.06                 &  0.13            & 0.38                                               \\
          r/socialism~\tiny\textbf{(trotskyists, khrushchev, gorbachev)}                     & 708                   & 0.13                         & 0.07                 &  0.16            & 0.28                                                 \\
          r/Anarchism~\tiny\textbf{(khrushchev)}                                             & 35                   & 0.17                         & 0.11                 &  0.16            & 0.00           \\
          \midrule
          \multicolumn{6}{c}{Stalinist Leaning \#2}  \\
          \midrule                                                                                                                                                                            
          Tankies~\tiny\textbf{(trots, de(-)stalinization, khrushchev, trot, trotskyists)}   & 4,044                   & 0.18                         & 0.08                 &  0.20            & 1.61                                                            \\
          r/socialism~\tiny\textbf{(stalin)}                                                  & 3,958                   & 0.16                         & 0.09                 &  0.17            & 0.73                                                 \\
          r/DemocraticSocialism~\tiny\textbf{(tankies)}                                       & 369                     & 0.13                         & 0.05                 &  0.19            & 1.63                                                   \\
          r/Anarchism~\tiny\textbf{(stalinists,ussr)}                                        & 2,393                   & 0.18                         & 0.09                 &  0.19            & 1.42           \\
          \midrule
          \multicolumn{6}{c}{Acceptance of Russian Narrative in Ukraine \#1}  \\                                         
          \midrule                                                                                                                                                                            
          Tankies~\tiny\textbf{(lpr,dpr)}                 & 627               & 0.15                         & 0.07                 &  0.17            & 0.96                                                            \\
          r/communism~\tiny\textbf{(donbass)}             & 36                  & 0.14                         & 0.10                 &  0.12            & 0.00                                               \\
          r/socialism~\tiny\textbf{(donetsk)}              & 65                  & 0.14                         & 0.07                 &  0.15            & 0.00                                                 \\
          r/Anarchism~\tiny\textbf{(donbass)}              & 48                  & 0.24                         & 0.17                 &  0.21            & 4.16                                                    \\
          \midrule
          \multicolumn{6}{c}{Acceptance of Russian Narrative in Ukraine \#2}  \\                                         
          \midrule                                                                                                                                                                            
          Tankies~\tiny\textbf{(zelensky)}                 & 1,687               & 0.23                         & 0.15                 &  0.22            & 2.54                                                            \\
        
          r/DemocraticSocialism~\tiny\textbf{(trump)}    & 15,053              & 0.08                         & 0.01                 &  0.15            & 0.61                                                   \\
          \midrule
          \multicolumn{6}{c}{Anti-Zionist Leaning }  \\ 
          \midrule                                                                                                                                                                            
          Tankies~\tiny\textbf{(zionist, zionists, zionism)}  & 1,933                  & 0.35                         & 0.29                 &  0.21            & 5.13                                                            \\
          Tankies~\tiny\textbf{(isntreal)}                  & 107                  & 0.22                         & 0.10                 &  0.23            & 2.83                                                            \\
          r/communism~\tiny\textbf{(israeli)}              & 95                     & 0.23                         & 0.20                 &  0.17            & 1.06                                               \\
          r/socialism~\tiny\textbf{(israel)}              & 2,501                  & 0.26                         & 0.20                 &  0.22            & 4.61                                                 \\
          r/DemocraticSocialism~\tiny\textbf{(jews)}    & 1,278                  & 0.12                         & 0.04                 &  0.17            & 0.62                                                   \\
          r/Anarchism~\tiny\textbf{(israel)}              & 1,229                  & 0.27                         & 0.22                 &  0.20            & 2.35                                                    \\
        
          \bottomrule    
      \end{tabular}
      
      \end{center}
  \end{scriptsize}
  \caption{Total number of posts containing keywords indicative of misalignment, as well as the mean and median severe toxicity scores, their standard deviations, and proportion of posts with high severe toxicity levels for each community.}
\label{tab:misalignment perspective}
  \end{table*}

\subsection{Validation}
\label{sec:appendix_misalignment_validation}

Just because two communities have misalignments and conceptual homomorphisms does not necessarily explain what those misalignments and homomorphisms mean, which is central to our goal of understanding tankies in relation to the far-left.

To address this, as mentioned in Section~\ref{sec:misalignment}, we perform a small-scale qualitative analysis utilizing a sample of posts containing misaligned/conceptually homomorphic word pairs to inform our interpretations. 
To further enhance the validity of our conclusions, we also implement a quantitative validation experiment using the \texttt{SEVERE\_TOXICITY} model provided by Perspective API. 
These experiments aimed to validate our interpretation of the misalignment of word pairs related to: 
1)~acceptance of Chinese Communist Party (CCP) narratives, 
2)~Stalinist leaning, 
3)~acceptance of Russian narratives in Ukraine, and 
4)~anti-Zionist leaning.

In our analyses, we group the misaligning/conceptually homomorphic word pairs of each community.
As an example, to validate our findings on the anti-Zionist leanings of tankies, we analyze the posts containing ``Zionist,'' ``isntreal,'' ``Zionists,'' and ``Zionism'' keywords for the tankies community, as well as the corresponding misaligning words for the other communities we analyze.
Nonetheless, in cases where keywords possess polarized or disparate meanings, we partition them for specific interpretations within certain communities (e.g., when validating the Stalinist leaning of tankies, we do not put ``Khrushchev'' and ``Stalin'' in the same keyword list).
Categories of our interpretations with the keywords used for each community can be seen from Table~\ref{tab:misalignment perspective}.

We do not validate our findings for toxic tone as we find tankies is the most toxic community among the communities we analyze (See Appendix~\ref{sec:perspective}).
Similarly, we do not validate findings related to tankies' leaning on state-sponsored media as our results reveal that tankies has the highest proportion of users sharing news outlets from ``Revisionist World Powers,'' which are generally known to be state-sponsored (See Appendix~\ref{sec:domain_main}).
We also do not perform further analysis regarding US politics, as the misalignment between tankies and r/DemocraticSocialism is the only substantial observation in our data.

\subsubsection{Qualitative Validation}
\label{sec:appendix_qualitative_validation}

Our analysis on the sampled posts (20 for each community and interpretation) suggests that tankies predominantly supports CCP, especially compared to the other far-left communities. 
We find many posts that show support for CCP and blames west for conducting anti-CCP propaganda -- e.g.:
\begin{quote}
  \textit{The media is compelled to target the CPC, which often leads to its misrepresentation.}

\end{quote}

During our analysis, we encounter posts that denies the Uyghur Genocide for each community with varying frequencies, while tankies primarily deny the occurrence of the genocide -- e.g.:

\begin{quote}
  \textit{To me, boarding schools serve as schools for potential terrorists, and China's approach seems more humane than the US's.}

\end{quote}

We also confirm that tankies mostly use ``rioters'' to define Hong Kong/Tiananmen Square protesters -- e.g.:

\begin{quote}
  \textit{HK rioters are frikkin psychos.}

\end{quote}

and,

\begin{quote}
  \textit{I won't dispute the Tiananmen Square massacre; rest in peace soldiers who died addressing the CIA-fueled unrest.}

\end{quote}

We find tankies talk negatively about Trotskyists and de-Stalinization --e.g.:

\begin{quote}
  \textit{Khrushchev was a big time idiot when it came to de-Stalinization.}

\end{quote}

Nevertheless, we observe mixed posts concerning Khrushchev.
This could be attributed to the fact that, although tankies predominantly oppose de-Stalinization and criticize Khrushchev on this basis, they also exhibit sympathy for him as a former leader of the USSR who presided over the Soviet invasion of Hungary.
We also find r/DemocraticSocialism use ``tankie'' as a pejorative term and show opposition to this community.

In our sample, we find numerous posts from tankies expressing support for the Russian invasion of Donbass -- e.g.:
\begin{quote}
  \textit{Safeguarding DPR and LPR from Nazis financed by NATO is a positive action.}

\end{quote}
In our analysis of the sampled posts from the other communities, we observe a mixture of content.
Notably, r/communism and r/socialism exhibit a higher frequency of posts expressing support for Russia compared to r/Anarchism.
Furthermore, our analysis reveals that the majority of the sampled posts from tankies mentioning Zelensky express opposition to him -- e.g.:

\begin{quote}
  \textit{Zelensky should've an azov symbol on his head.  }

\end{quote}

We also see the sampled posts of tankies are predominantly anti-Zionist -- e.g.:
\begin{quote}
  \textit{Zionism equates to Fascism.}

\end{quote}
and,

\begin{quote}
  \textit{That's Isntreal and nah, we don't endorse Israel.}

\end{quote}
We observe numerous posts expressing opposition to Israel or Israeli policies across other communities. 
However, r/DemocraticSocialism exhibits the lowest frequency of such posts compared to the other communities.

\subsubsection{Quantative Validation}
\label{sec:appendix_quantitative_validation}

To quantitatively validate our interpretations, we first leverage the \texttt{SEVERE\_TOXICITY} model of the Perspective API.
We then compare the high toxicity levels of various misaligning keywords for each interpretation.
Table~\ref{tab:misalignment perspective} shows the basic statistics and proprotions of high \texttt{SEVERE\_TOXICITY} scores for the keywords indicative of misalignment.

\descr{Acceptance of CCP Narrative.}
Our analysis reveals that tankies exhibit the lowest proportion of high \texttt{SEVERE\_TOXICITY} for posts containing the keyword ``CPC'' when compared to the posts of other communities containing the keyword ``CCP.''
Conversely, tankies display a higher proportion of high scores for the keywords related to Uyghur Genocide. 
Based on our sampled data, we infer that tankies tend to adopt a more toxic tone when denying the Uyghur Genocide, while other communities engage in less toxic discourse when discussing Uyghur-related topics.
Furthermore, we observe that r/communism and r/socialism demonstrate a reduced fraction of high toxicity when addressing ``camps,'' which are mostly related to concentration camps established the by the CCP or Nazis.
Finally, we find tankies have higher proportion of high scores when mentioning ``rioters'' (which are possibly towards Hong Kong protesters) compared to other communities' posts mentioning ``protesters/protestors.'' 
Additionally, tankies have the second highest proportion of high \texttt{SEVERE\_TOXICITY} scores for their posts containing Tiananmen Square compared to the other communities posts contain ``riots.''

\descr{Stalinist Leaning.}
We find that tankies display the highest proportion of high \texttt{SEVERE\_TOXICITY} scores for posts featuring anti-Stalinist or de-Stalinization-related keywords when compared to r/communism, r/socialism, and r/Anarchism.
Furthermore, we observe more comparable proportions when comparing the keywords used by tankies with the Stalinist or pro-USSR keywords present in r/socialism and r/Anarchism.
In addition, we identify a similar proportion of high scores in r/DemocraticSocialism's usage of tankies, further corroborating the negative sentiment of tankies towards anti-Stalinism and de-Stalinization. 

\descr{Acceptance of Russian Narrative on Ukraine.}
We find tankies exhibit the second highest proportion of high \texttt{SEVERE\hspace{0pt}\_TOXICITY} scores for posts with keywords related to Ukraine's Donbass region.
Nevertheless, less than 1\% of tankie posts display high scores, which may be attributed to their support for the self-proclaimed republics in Donbass. 
Moreover, tankies have more than four times higher proportion of high scores for posts containing ``Zelensky'' compared to r/DemocraticSocialism's posts containing ``Trump.''
This finding serves as an additional indicator of tankies' alignment with the Russian narrative. 

\descr{Anti-Zionist Leaning.}
Our analysis reveals that over 5\% of tankie posts containing Zionism-related keywords exhibit high \texttt{SEVERE\_TOXICITY} scores, which is the highest proportion compared to other communities.
Intriguingly, tankies display an even greater proportion of elevated scores for posts featuring Zionism-related keywords than for their own usage of the term ``isntreal.''

\begin{figure*}[th!]
  \centering
  \includegraphics[width=1\textwidth]{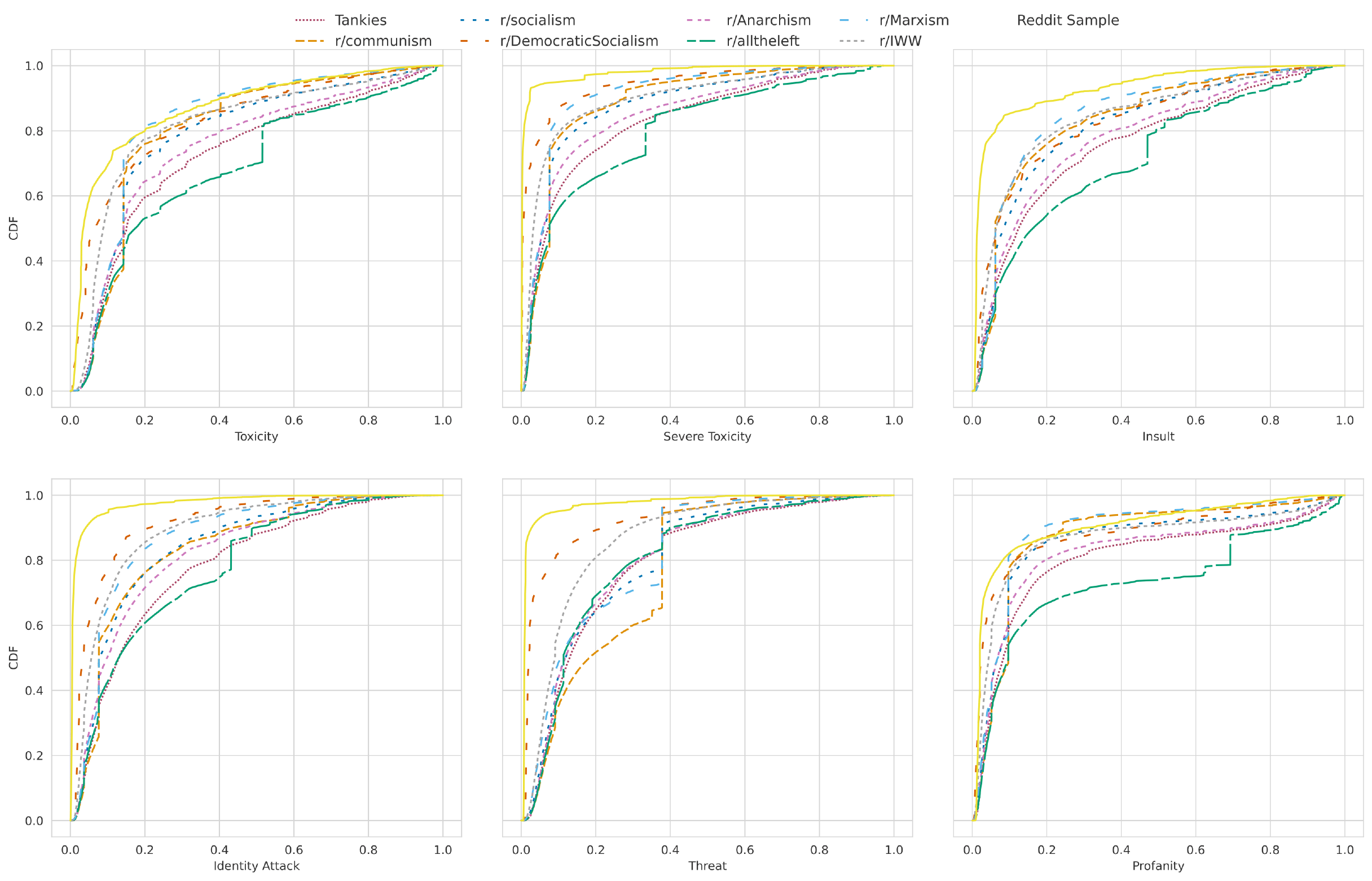}
  \caption{CDF of Perspective API attributes of the far-left cluster. Tankies have the highest proportion of high scores for identity attack and threat among far-left communities and the second-highest for toxicity, severe toxicity, profanity, and insult behind r/alltheleft.}
  \label{fig:perspective_all}
\end{figure*}

\begin{table*}[t]
  \centering
  \begin{adjustbox}{width=\textwidth}
    \begin{tabular}{l|rrrrrr}
        \toprule
        \multicolumn{1}{c}{Community}                             & \multicolumn{1}{c}{Toxicity}    &  \multicolumn{1}{c}{Severe Toxicity}         & \multicolumn{1}{c}{Identity Attack}               & \multicolumn{1}{c}{Profanity}                     & \multicolumn{1}{c}{Threat}                    & \multicolumn{1}{c}{Insult}\\
        
        \midrule
        Tankies                                                   &8.02                        & 1.90         & 2.00    & 9.03             & 2.10       & 4.90    \\
        r/communism                                               &2.53                        & 0.57         & 0.60    & 3.15             & 0.65       & 1.74  \\
        r/socialism                                               &4.41                        & 1.00         & 0.93    & 5.59             & 1.26       & 2.51  \\
        r/DemocraticSocialism                                     &2.64                        & 0.34         & 0.22    & 2.45             & 0.26       & 1.42  \\
        r/Anarchism                                               &6.63                        & 1.65         & 1.38    & 8.48             & 1.81       & 3.73  \\
        r/alltheleft                                              &9.63                        & 3.81         & 1.42    & 10.74            & 1.69       & 6.92   \\
        r/Marxism                                                 &1.98                        & 0.46         & 0.36    & 2.47             & 0.36       & 1.66  \\
        r/IWW                                                     & 4.62                       & 0.99         & 0.37    & 6.03             & 0.70       & 2.62  \\
        Reddit Sample                                             & 1.65                       & 0.04         & 0.04    & 1.70             & 0.02       & 0.23 \\
        \midrule
        Mean (Far-Left except Tankies)                            & 4.63                      &1.26           &0.75     & 5.56             & 0.96       & 2.94  \\
        \bottomrule
    \end{tabular}
  \end{adjustbox}
    \caption{ Fractions of high perspective scores for each far-left community. We also include the high perspective score frequencies for the Reddit sample for comparison.}
    \label{tab:far_left_high_perspectives}
\end{table*}

\section{Toxicity Analysis}
\label{sec:perspective}

In Section~\ref{sec:misalignment}, our findings reveal that tankies tend to refer to other ideologies with pejorative terms, signaling their toxic behavior.
To further assess the tone of tankies' conversations and compare their online behavior with other far-left communities, we perform a set of experiments using Perspective API models.

\descr{Perspective API.}
The Perspective API~\cite{jigsaw2018perspective} is a widely used~\cite{aliapoulios2021gospel,horta2021platform} tool for measuring toxicity.
Although it has limitations, e.g., there are issues of bias and questions of performance when encountering conversation patterns that it was not trained on, at scale it provides a decent measure for comparison between online communities. 
The API provides six production models: 
1)~\texttt{TOXICITY}, 
2)~\texttt{SEVERE\_TOXICITY}, 
3)~\texttt{INSULT}, 
4)~\texttt{IDENTITY\_ATTACK}, 
5)~\texttt{THREAT}, and 
6)~\texttt{PROFANITY} (See~\cite{perspective} for full details on the models).
We consider a threshold of 0.8, defined as ``high'' for the SEVERE\_TOXICITY scores by Hoseini et al.~\cite{hoseini2023globalization}.
To have a baseline for the comparisons, we sample 0.5\% of the Reddit posts during the dataset's timeline, which accounts to more than 36 M posts.

\begin{table*}[ht!]
  \centering

  \begin{adjustbox}{width=\textwidth}
    \begin{tabular}{lrrlrrlrr}
        \toprule
        \multicolumn{3}{c}{Toxicity}                             &  \multicolumn{3}{c}{Severe Toxicity}         & \multicolumn{3}{c}{Identity Attack}               \\
        \midrule
        Named Entity     & \%High & \%Posts                &Named Entity    & \%High & \%Posts      &Named Entity     & \%High & \%Posts          \\    
        \midrule
        Stupidpol        & 51.93       & 0.017                   & Facist         & 8.06    & 0.008             & Muslims       & 41.80  & 0.232                   \\
        Mayo             & 23.12       & 0.011                   & Yankee         & 7.78    & 0.017             & Jews          & 34.03  & 0.138                    \\
        Ceausescu        & 23.00       & 0.008                   & Stupidpol      & 7.72    & 0.017             & Muslim        & 28.90  & 0.210                    \\
        Amerikkkans      & 18.86       & 0.007                   & Amerikkka      & 7.49    & 0.122             & Asians        & 28.20  & 0.071                    \\
        Jack             & 17.52       & 0.033                   & Mayo           & 7.48    & 0.011             & Arabs         & 25.0   & 0.044                    \\
        John Oliver      & 17.41       & 0.016                   & Blinken        & 6.86    & 0.007             & Islam         & 24.59  & 0.090                    \\
        Yankee           & 17.21       & 0.017                   & Yankees        & 6.45    & 0.011             & Hindus        & 24.19  & 0.008                    \\
        Elon             & 16.56       & 0.011                   & Erdogan        & 6.42    & 0.018             & Mexicans      & 24.01  & 0.014                    \\
        Kyle             & 16.54       & 0.009                   & Michael        & 6.34    & 0.018             & Africans      & 23.73  & 0.034                    \\
        ACAB             & 16.16       & 0.015                   & Chechens       & 5.82    & 0.013             & White         & 23.42  & 0.008                    \\
        \bottomrule
        \toprule
          \multicolumn{3}{c}{Profanity}                     & \multicolumn{3}{c}{Threat}                    & \multicolumn{3}{c}{Insult}\\
        \midrule
        Named Entity     & \%High & \%Posts          &Named Entity     & \%High & \%Posts      &Named Entity     & \%High & \%Posts \\
        \midrule
        Charlotte      & 33.33   & 0.009                  & Ceausescu  & 15.92  & 0.007                   & Stupidpol         & 66.09     & 0.014                                                           \\
        Stupidpol      & 27.46   & 0.017                  & Minecraft  & 14.54  & 0.048                   & Ceausescu         & 22.12     & 0.008                                                           \\
        Jack           & 26.92   & 0.033                  & 6 Million  & 14.37  & 0.012                   & Four Years        & 19.00     & 0.008                               \\
        24 hours       & 26.08   & 0.008                  & Chechens   & 13.22  & 0.013                   & Daniel Dumbrill   & 17.18     & 0.009                                    \\
        Mayo           & 23.80   & 0.011                  & 24 Hours   & 12.17  & 0.008                   & Mayo              & 17.00     & 0.008                                                    \\
        Qanon          & 21.39   & 0.015                  & Rosa       & 12.08  & 0.020                   & Joe Rogan         & 17.00     & 0.010                            \\
        Brasil         & 20.51   & 0.015                  & Inshallah  & 10.06  & 0.057                   & Facist            & 16.93     & 0.008                                    \\
        Alex Jones     & 20.46   & 0.012                  & Iraqis     & 9.86   & 0.022                   & Serpentza         & 16.80     & 0.009          \\
        All day        & 20.35   & 0.032                  & Romanovs   & 9.67   & 0.008                   & Blinken           & 16.66     & 0.007          \\
        Elon           & 20.24   & 0.011                  & Kulaks     & 9.38   & 0.017                   & Amerikkkans       & 16.03     & 0.007        \\

        \bottomrule
    \end{tabular}
  
  \end{adjustbox}
    \caption{ Top 10 Named Entities of tankies that appeared more than 100 times with high perspective scores with their high score frequencies and proportions in total number of posts of tankies.}
    \label{tab:tankies_ner_perspective}
\end{table*}

\begin{figure*}[th!]
  \centering
  \includegraphics[width=1\textwidth]{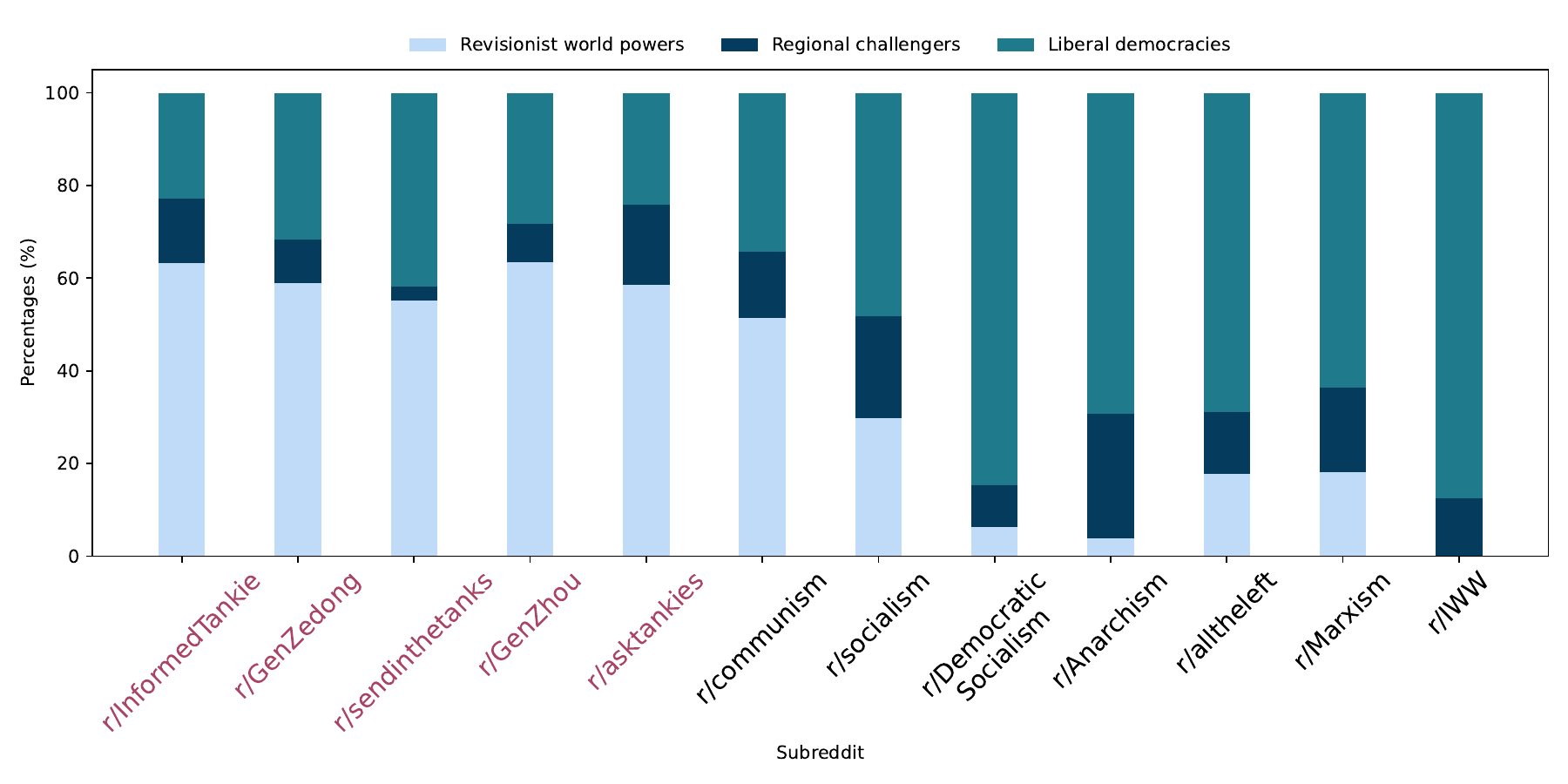}
  \caption{Proportions of sharing amount of news outlets from Revisionist World Powers, Regional Challengers, and Liberal Democracies by each far-left community we analyze. Tankie subreddits are specified with red. We find that tankies have the highest proportion for sharing new outlets from ``Revisionist Super Powers.''}
  \label{fig:stacked_bar}
\end{figure*}

\descr{Results.}
Figure~\ref{fig:perspective_all} shows the cumulative distribution functions (CDFs) for each model within the far-left cluster.
Our analysis reveals that tankies tend to have higher scores than other far-left communities (excluding r/alltheleft) for all Perspective API models. 
Additionally, all far-left communities have higher Perspective API scores than the baseline Reddit sample.

Specifically, tankies have the highest proportion for both scores $\geq 0.5$ and high scores (i.e., scores $ \geq 0.8$) for \texttt{IDENTITY\hspace{0pt}\_ATTACK} and \texttt{THREAT}, and second highest for the remaining models. 
Table~\ref{tab:far_left_high_perspectives} shows that tankies have nearly twice as many high scores than the mean of other far-left communities.

We confirm that the score distributions for each model are significantly different between tankies and other far-left communities using a 2-sample KS test  ($p < 0.01$ for all after adjustment for multiple testing using the Benjamini-Hochberg method). 
These results indicate that tankies tend to make posts with higher levels of toxicity, insult and profanity compared to other far-left communities, excluding r/alltheleft.
Tankies also tend to make posts with more identity attacks and threats than all other far-left communities.

Next, we examine the named entities in tankies' posts that have high Perspective API scores by removing any named entities that appear fewer than 100 times.
In Table~\ref{tab:tankies_ner_perspective}, we present the top 10 named entities ranked by the fraction of posts that mention the entity and score high across all Perspective models.
For all models except \texttt{IDENTITY\_ATTACK}, the most commonly mentioned named entities are primarily related to the US (e.g., Amerikkkans, Yankee, Charlotte, Qanon), public figures and politicians (e.g., John Oliver, Elon Musk, Kyle Rittenhouse, Anthony Blinken, Erdogan, Alex Jones, Joe Rogan), or countries/nationalities (e.g., Chechens, Brasil, Iraqis).
For the \texttt{IDENTITY\_ATTACK} model, the most frequently mentioned named entities are typically religious or ethnic groups. 
Tankies appear to primarily target Muslims and Jews, with 41.80\% and 34.03\% of the posts mentioning these groups having high \texttt{IDENTITY\_ATTACK} scores; the highest proportion of high \texttt{IDENTITY\_ATTACK} scores for Muslims and Jews of any far-left community we analyze.
Furthermore, we observe that tankies attack the identities of Asians, Arabs, Hindus, Mexicans, Africans, and Whites in more than 20\% posts mentioning these identities.

\descr{Takeaways.}
Our analysis shows that tankies have the highest proportion of high scores for \texttt{IDENTITY\_ATTACK} and \texttt{THREAT} among far-left communities, and they have the second highest proportion of high scores for \texttt{TOXICITY}, \texttt{SEVERE\_TOXICITY}, \texttt{PROFANITY}, and \texttt{INSULT}, behind r/alltheleft.
Although our findings indicate that US politics are not the primary focus for tankies, they still express strong opinions about US-related events, conspiracy theories, politicians, and public figures.
Finally, we observe that tankies frequently target Muslims and Jews in their posts.
  
\section{Domain Analysis}
\label{sec:domain_main}

In Section~\ref{sec:misalignment}, we find that tankies tend to use Chinese and Russian news outlets in a similar way to how r/\hspace{0pt}DemocraticSocialism uses US news outlets.
Based on this discovery, we compare the sharing of news outlets by far-left communities according to ``state types''. 
We use a set of selected outlets to represent three geopolitical groups identified by Xu et al.\cite{xu2022nationalizing}:

\begin{enumerate}
    \item \emph{Revisionist World Powers: } 17 different state-sponsored or state-controlled news outlets from China and Russia.
    \item \emph{Regional Challengers: } 6 different news outlets from Saudi Arabia, Qatar, Iran, Venezuela, and Turkey.
    \item \emph{Liberal Democracies: } 9 different news outlets from Korea, UK, USA, Germany, EU, France, Israel, and Japan.
   
  \end{enumerate}

Figure~\ref{fig:stacked_bar} shows the distribution of the three geopolitical groups among tankie subreddits and other far-left communities. 
On average, tankies share outlets from ``Revisionist World Powers'' 60\% of the time, while r/communism shares them 51.3\% of the time.
Other far-left communities, in order of decreasing sharing frequency, are r/socialism (29.7\%), r/Marxism (18.1\%), r/alltheleft (17.8\%), r/DemocraticSocialism (6.3\%), and r/Anarchism (3.7\%), with r/IWW not sharing any.
While tankies are the top far-left community in terms of sharing outlets from ``Revisionist World Powers,'' they rank last in terms of sharing outlets from ``Regional Challengers'' and ``Liberal Democracies,'' with averages of 10\% and 30\%, respectively.

\descr{Takeaways.}
We examine the domains that reflect the characteristics of tankies and compared their sharing of news outlets with other far-left subreddits. 
We find that tankies have a higher proportion of shared news outlets from ``Revisionist World Powers,'' which comprise state-sponsored or state-controlled outlets from China and Russia, compared to their similar ideological counterparts. 
In contrast, they have the lowest proportion of shared news from ``Regional Challengers'' and ``Liberal Democracies.''
Given the history of misinformation and disinformation from news outlets in Russia and China~\cite{bandurski2022}, these findings may suggest that tankies prefer sources of misleading information compared to other far-left communities.

\end{document}